\documentclass[12pt]{aastex}			

\usepackage{graphicx}	 			

\graphicspath{{./fig/}}  

\newcommand{\matvec}[1]{\ensuremath{\mathbf{#1}}}

\title{Reducing the Dimensionality of Data: Locally Linear Embedding of Sloan Galaxy Spectra}	

\author{Jake Vanderplas, Andrew Connolly\\
 University of Washington}

\date{\today}				

\begin{document}			

\begin{abstract}
We introduce Locally Linear Embedding (LLE) to the astronomical community 
as a new classification technique, using SDSS spectra as an example data set. 
LLE is a nonlinear dimensionality reduction technique which has been studied 
in the context of computer perception.  We compare the performance of LLE to 
well-known spectral classification techniques, e.g.\ principal component 
analysis and line-ratio diagnostics. We find that LLE combines the strengths 
of both methods in a single, coherent technique, and leads to improved 
classification of emission-line spectra at a relatively small computational 
cost.  We also present a data subsampling technique that preserves local 
information content, and proves effective for creating small, efficient 
training samples from a large, high-dimensional data sets. Software used 
in this LLE-based classification is made available.
\end{abstract}

\bibliographystyle{apj}
\maketitle

\section{Introduction}

With the development of large-scale, astronomical photometric and
spectroscopic surveys such as 
the Sloan Digital Sky Survey \citep[SDSS;][]{York00}, 
the Two Micron Survey \citep[2MASS;][]{Skrutskie06}, 
the Panoramic Survey Telescope 
and Rapid Response System \citep[PanSTARRS;][]{Kaiser02},
and the Canada France Hawaii Telescope Legacy Survey (CFHTLS),
the amount of data available to astronomers has 
increased dramatically. While these data sets enable much new science,
we face the question of how we extract information efficiently from
data sets that are many terabytes in size and that contain a large
number of measured attributes. In other words, how do we classify 
objects in a physically meaningful way when the data span a large number 
of dimensions and we do not know {\it a priori}  which of those 
dimensions are important?

The question of the classification of astronomical data remains
open. We can, however, draw upon a number of successful classification
schemes, such as Hubble's morphological classification of galaxies
and Morgan's spectral classification of stars, and note the attributes
they have in common. First, classifications
must be simple: Hubble's initial classification into seven
different types remains the dominant scheme used in astronomy despite
numerous extensions based on subtypes
\citep[e.g.][]{deVaucouleurs59,VanDenBergh76}, concentration indices and
asymmetry parameters \citep{Abraham94}, clumpiness
\citep{Conselice03}, Gini coefficients \citep{Abraham03}, and neural
networks \citep{Lahav96}.  Second, they must be physically meaningful:
the Hubble morphological scheme relates to the dynamical history of
the galaxy. Finally, the number of classes must account for
the intrinsic uncertainty in the properties of the observed sources
(i.e.\ we do not want to over-fit the data if the scatter in the
classifications is large).

These principles can be applied to spectral data, which have the
advantage of being easier to map to physical properties of sources
than images.
Continuum shapes, emission line ratios, and absorption indices can all be
used as proxies for star formation rate, stellar age, dust absorption
and metallicity \citep{Heavens2000,Reichardt01}.  
As such, spectral classification, and in 
particular statistical techniques for spectral classification, is
more broadly used than morphological classification. Examples of
this include classifications based on color \citep{Cramer79,Kunzli97}, 
spectral indices \citep[e.g.\ Lick indices;][]{Gulati99}, 
model fitting \citep{BruzualCharlot03}, and more general 
non-parametric techniques \citep{Connolly95,Folkes96,Yip04,
Richards09}.

The difficulty that spectral classification poses comes from the
inherent size and dimensionality of the data. For example, the SDSS
spectroscopic survey comprises over 1.5 million spectra, 
each with 4000 spectral 
elements covering 3800\AA\ to 9800\AA. Providing a simple and physical
classification requires that we reduce the dimensionality of the data in a
way that preserves as many of the physical correlations as
possible. A number of techniques have been developed.
One such approach that has met with success is the
Karhunen-Loeve (KL) Transform or Principal Component Analysis (PCA). PCA is a
linear projection of data onto a set of orthogonal basis functions
(eigenvectors). It has been applied to galaxy spectra \citep{Connolly95,
  Folkes96, Sodre97, Bromley98, Galaz98, Ronen99, Folkes99, Yip04, Budavari09},
QSOs \citep{Francis92,Boroson92,Yip04a}, 
stellar spectra \citep{Singh98, Bailer-Jones98} as well as to measured
attributes of stars and galaxies \citep{Efstathiou84}. The PCA projection
for galaxy spectra has been shown to correlate with star formation rate 
\citep{Connolly95,Madgwick-Somerville-03}  
and is now actively used as a classification scheme in the
SDSS and 2dF spectroscopic surveys.

As we will discuss within this paper, the linearity of the KL
transform, while providing eigenspectra that are relatively simple to
interpret, has an underlying weakness. It cannot easily (nor compactly)
express inherently non-linear relations within the data (such as dust
obscuration or the variation in spectral line widths). Spectra that
have a broad range in line-widths require a large number of
eignespectra to capture their intrinsic variance which often results
in the continuum emission and emission lines being treated
independently \citep{Gyory08}. In this paper we consider a new approach to
classification, using a dimensionality reduction technique that preserves the
local structure within a data set: Local Linear Embedding 
\citep[LLE;][]{Roweis2000}. 
In Section \ref{dimensionality_reduction} we will discuss the properties 
of spectral classification using Principal Component Analysis and line ratios. 
In Sections \ref{LLE_intro} and \ref{LLE_on_spectra}
 we introduce LLE and apply it to the
classification of spectra from the SDSS. 
In Section \ref{LLE_considerations} we consider some of the practical
details to consider when applying the LLE procedure.  
In Section \ref{discussion} we discuss the utility of the
LLE as a dimensionality reduction scheme as well as its
performance relative to classical spectral classification techniques.

\section{Dimensionality Reduction and Classification of  Spectral Data}
\label{dimensionality_reduction}

\subsection{Principal Component Analysis}

Over the last decade PCA or KL classification has been applied to a
wide range of spectral classification problems. From non-parametric
classification \citep{Connolly95} to extracting star formation properties 
\citep{Ferreras06} to
identifying supernovae within SDSS spectra 
\citep[][Krughoff et al. in prep]{Madgwick03a},
the many applications of PCA have demonstrated
the importance of dimensionality reduction in classification. This
technique seeks to decompose a dataset into a linear combination of a
small number of principal components, such that each principal
component describes the maximum possible variance.
\begin{equation}
  S( \lambda ) = \sum_i a_i e_i( \lambda ) 
\end{equation}
where $S(\lambda)$ is the spectrum to be projected, $a_i$ are the expansion 
coefficients, and $e_i(\lambda)$ the orthogonal eigenspectra.

Schematically, PCA is equivalent to finding a best-fit linear subspace
to the entire dataset, such that the variance of the data projected
into this subspace is maximized.  \citet{Yip04} present a robust PCA
approach to spectral classification of the SDSS spectra.  In it, they
find that the information within the SDSS main galaxy sample can be
almost completely characterized by the first dozen eigenmodes.
This represents a reduction in data size by a factor of nearly 400
over the complete spectra. Furthermore, the coefficients 
of the first three eigencomponents correlate
with physical properties of the galaxies such as star-formation rate,
and post star-burst activity. In fact, PCA decomposition can lead to a
single parameter description of galaxy spectra: the mixing angle
$\phi$ between the first and second eigencoefficients.  This mixing
angle has been shown to correlate well with galaxy spectral type
\citep{Connolly95, 2df} and can be used for an approximate separation
between quiescent and emission-line galaxies.

For continuum emission PCA has a proven record in representing the
variation in the spectral properties of galaxies. It does not,
however, perform well when reconstructing high-frequency structure
within a spectrum (i.e.\ the distribution of emission lines, lines-ratios,
 and line-widths). This is due to two effects: principal
components preserve the total variance of the system, i.e.\ the overall
color of the spectrum, and principal components express linear
relations between components. High frequency, or local, features do
not contribute significantly to the total variance and are, therefore,
not represented in the primary eigencomponents. Variations in 
line-widths and, often, line-ratios are also inherently non-linear in nature
(due to the impact of dust or variations in the galaxy mass) and
are not compactly represented as a linear combination of orthogonal
components. This is shown in \citet{Yip04} where the majority of
galaxies can be represented by three eigenspectra but emission-line
galaxies require up to eleven components to express their
variance. PCA performs poorly when distinguishing between
emission line galaxies, e.g.\ separating broad-line QSOs from
narrow-line QSOs or star-forming galaxies.

\subsection{Line-Ratio Diagrams}

Line-ratio diagrams, sometimes known as \textit{BPT plots}
\citep{BPT81} or \textit{Osterbrock diagrams} \citep{Osterbrock85},
are one answer to the problems associated with PCA classification.
Where PCA classifies based on the total variance within a spectrum,
which weights the continuum emission more than individual emission
lines, line-ratio diagrams are sensitive to emission-line
characteristics while ignoring continuum properties.  This makes them
appropriate for diagnostics that distinguish between galaxies with
narrow emission lines.  Line-ratio diagrams, however, do not account
for continuum emission that might provide additional information on the 
physical state of a galaxy.  They are ill-suited
to classify non-emitting galaxies and galaxies with low signal-to-noise for
individual lines, and fail completely for galaxies with certain types of
emission, e.g.\ broad-line QSOs (often defined as those with 
emission line FWHM larger than $\sim 1200$ km s$^{-1}$). Line-ratio diagrams,
then, are useful for the classification of only a small subset of observed 
objects.

\subsection{A Joint Approach to Spectral Classification}

Clearly, PCA and line-ratio diagrams serve complimentary purposes.
PCA takes into account broad, low-frequency features, and can
distinguish galaxies based on their continuum properties.  Line-ratio
diagrams, by design, depend only on the detailed features of emission
lines, and therefore distinguish spectra based on their narrow,
high-frequency features. If both types of information could be combined into a
joint classification scheme, one that treats the spectrum as a whole
rather than trying to isolate individual features, more insight
may be gained into the physical state of a given object. 
This would be particularly helpful for automated classification within
spectroscopic surveys, where a single, coherent technique could be 
used to identify objects which could then be analyzed further by
specialized pipelines.  The requirements of such a
general classification are: it must be sensitive to both low and high
frequency spectral information, it must be able to account for
non-linear relationships between the spectral properties, it should be
robust to outliers within the data and it should be able to express
the properties of a galaxy in terms of a small number of physically
motivated parameters. In this paper we consider Local Linear Embedding
as a solution to these questions.

\section{Locally Linear Embedding}
\label{LLE_intro}

Locally Linear Embedding (LLE) is a nonlinear dimensionality reduction
technique that seeks to find a low-dimensional projection of a higher
dimensional dataset which best preserves the geometry of local
neighborhoods within the data \citep{Roweis2000}.  It can be thought
of as a non-linear generalization of PCA: rather than projecting onto
a global linear hyper-surface, the projection is to an arbitrary
nonlinear surface constrained locally within the overall parameter
space (see Figure \ref{LLE-ex}). In this way it is superficially similar
to, e.g.\ Non-Linear PCA and Principal Curves/Manifolds \citep[see][for an 
introduction to and astronomical application of 
these techniques]{Einbeck07}.  These 
generalizations seek a lower-dimensional projection which optimally 
represents the information contained in each point, 
such that the original data can
be reconstructed from the projection.  LLE, on the other hand, seeks a
projection which optimally represents the relationship between nearby points.
Accurate reconstruction of data is possible to an extent (see section 
\ref{new_points}), but is not the primary goal of an LLE projection.

The LLE algorithm consists of two parts: first, a set of local 
weights is found which parametrize
the linear reconstruction of each point from its nearest neighbors; 
second, a global minimization is performed to determine a lower-dimensional 
analog of the dataset which best preserves these local 
reconstruction weights.  

\subsection{Description of the LLE Algorithm}

For notational conventions, refer to Appendix \ref{notation}.  We
initially consider a series of points in a high dimensional space
$\matvec{x_i}$ (in the case of SDSS, each spectrum is represented by point 
in a $D_{in} = 4000$ dimensional space) 
such that the set of these points, $\matvec{X}$, is given by
$\matvec{X} = [\matvec{x_0},\matvec{x_1},\matvec{x_2},
... \matvec{x_N}],\ \matvec{x_i} \in \mathbb{R}^{D_{in}}$. We map this
to a lower dimensional space $\matvec{Y} =
[\matvec{y_0},\matvec{y_1},\matvec{y_2},...\matvec{y_N}],\
\matvec{y_i} \in \mathbb{R}^{D_{out}}$, with $D_{in} > D_{out}$. For
each point \matvec{x_i}, let $\matvec{n}^{(i)} =
[n_1^{(i)},n_2^{(i)},...n_K^{(i)}]^T$ be the indices of the nearest
neighbors, such that $\matvec{x}_{n_j^{(i)}}$ is the $j$th nearest
neighbor of \matvec{x_i}.  Also, let $\matvec{w}^{(i)} =
[w_1^{(i)},w_2^{(i)},...w_K^{(i)}]^T$ represent the reconstruction
weights associated
with the $K$ nearest neighbors of \matvec{x_i}.\footnote{Nearest
  neighbors can be defined based on any general distance metric.  In
  this work, Euclidean distance will be used unless otherwise noted.}

The key assumption is that each point \matvec{x_i} lies near a locally-linear,
low-dimensional manifold, such that a tangent 
hyperplane is locally a good fit. If this is the case,
a point can be accurately represented as a linear combination of its
$K$ nearest neighbors, with $K \ge D_{out}$.
The error in reconstruction can be thought of as the distance from
this manifold to the point in question. A convenient form of this
error is the reconstruction cost function
\begin{equation}
 \label{weight_E}
 \mathcal{E}_1^{(i)}(\matvec{w}^{(i)}) = \Bigg| \matvec{x_i} - \sum_{j=1}^K w_j^{(i)}\matvec{x}_{ n_j^{(i)} }\Bigg|^2.
\end{equation}
By minimizing $\mathcal{E}_1^{(i)}(\matvec{w}^{(i)})$ subject to the constraint
\begin{equation}
 \sum_j w_j^{(i)} = 1 ,
\end{equation} 
we find the optimized local mapping.  This local mapping has two
important properties: first, it is invariant under scaling,
translation, and rotation; second, it encodes the relationship
between points of the local neighborhood in a way that is agnostic to
the dimensionality of the parameter space, i.e., it equally well
encodes the properties of the neighborhood in observational parameter
space, as well as the properties of the neighborhood as 
projected onto the
local tangent hyperplane.  These two facts lie at the core of the LLE
algorithm.

Once the weights $\matvec{w}^{(i)}$ are determined for each point 
$\matvec{x_i}$, the same weights are used to determine the projected 
vectors \matvec{y_i} which minimize the global cost function,
\begin{equation}
 \label{proj_E}
 \mathcal{E}_2(\matvec{Y}) = \sum_{i=1}^N \Bigg| \matvec{y_i} - \sum_{j=1}^K w_j^{(i)}\matvec{y}_{n_j^{(i)}}\Bigg|.
\end{equation}
Notice the symmetry inherent in equations \ref{weight_E} and \ref{proj_E}.  
Intuitively, the fact that the input data \matvec{X} and the projected data 
\matvec{Y} optimize these linear forms shows that the local 
neighborhoods of \matvec{X} and \matvec{Y} have similar properties.

Computationally, each of these steps can be implemented efficiently
using optimized linear algebra methods \citep[See, for
example,][]{Roweis2000,deRidder2002}.  We discuss algorithmic
considerations in the Appendices \ref{finding_w}, \ref{finding_y} and
\ref{algorithm}.

\subsection{Dimensionality Reduction: A Simple Example}
\label{simple-example}
\begin{figure}
 \centering
 \includegraphics[height=0.8\textheight]{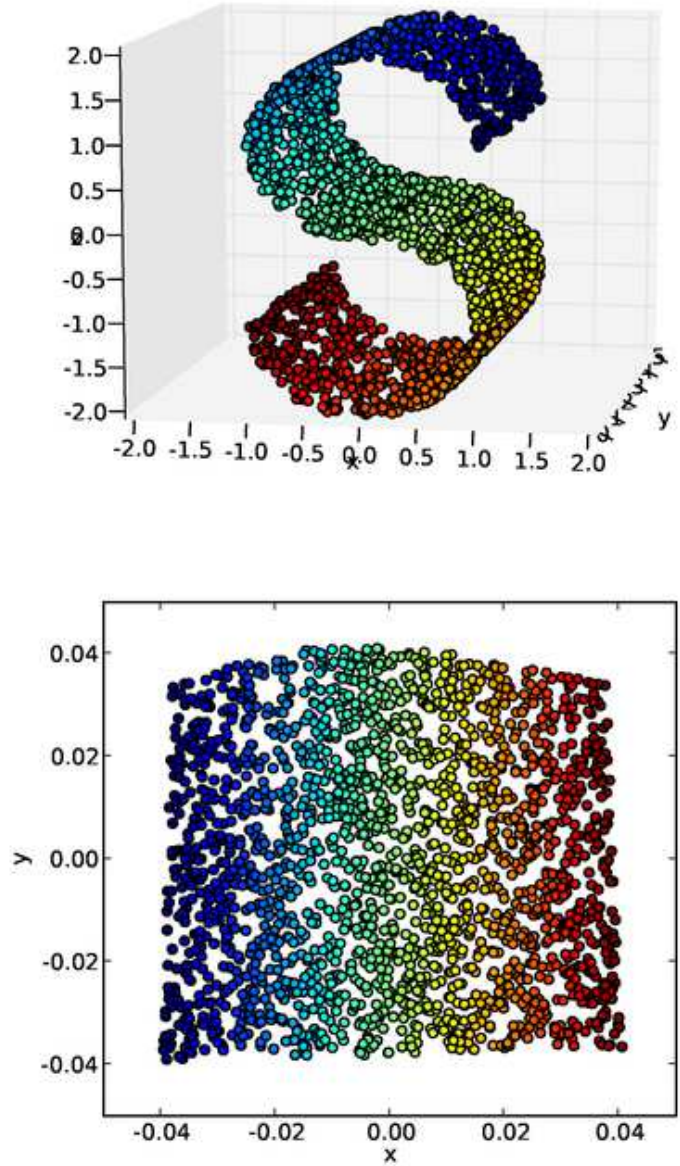}
 \caption{The canonical ``S-curve'' test set for nonlinear data-reduction.
   \textit{top:} $N=2000$ points randomly sampled from a bounded 
   2-manifold (eqn. \ref{S-def}) embedded in 3-space.  Points are colored
   to highlight the connection of local neighborhoods.  \textit{bottom:} 
   the 2 dimensional LLE projection of this manifold, with $K=15$.  Note 
   that points which are clustered in the 3-dimensional parameter space
   are also clustered in the 2-dimensional embedding space.  This shows 
   that LLE reconstructs the desired embedding.}
 \label{LLE-ex}
\end{figure}

We provide a simple and classical example of the performance of the
algorithm in Figure \ref{LLE-ex}.  The first plot shows a particularly
simple nonlinear test set: the ``S-curve'', 
with $N=2000$ data points
in three dimensions uniformly selected from the two-dimensional
bounded manifold described by
\begin{equation}
  \label{S-def}
  \begin{array}{ll}
    \left.
    \begin{array}{l}
      x = \sin(\theta)\\
      z = \frac{\theta}{|\theta|}(\cos(\theta)-1))\\
    \end{array} \right \} 
    & -1.5\pi\le\theta\le 1.5\pi \\
    0<y<5 & \\
  \end{array}
\end{equation}
A linear dimensionality reduction such as PCA would not recover the
inherently nonlinear shape of the two-dimensional manifold: it would find
three principal axes within the data.  An ideal
nonlinear reduction would, in effect, unwrap this manifold, and map it
to a flat plane which preserves local clustering -- i.e.\ the
colors would remain unmixed.

The second plot in Figure \ref{LLE-ex} shows the two dimensional 
LLE projection applied to this data, with $K=15$.  It is clear that LLE, in
this case, successfully recovers the desired 2-dimensional
embedding. Points of similar colors (which are clustered within the
higher dimensional space) are clustered in the projected space. As we
will discuss in Section \ref{choosing_parameters}, 
the effectiveness of this projection will
depend on the size of the region that we consider local (i.e.\ the
number of nearest neighbors) as well as the assumed dimensionality
of the embedded manifold.

The above test data densely sample a simply-connected manifold.  In the case
of sparsely sampled, or non-connected manifolds (e.g.\ due to missing data,
survey detection limits, etc.), LLE is less successful at
recovering the lower-dimensional projection.  This issue can be addressed
using various modifications to the algorithm 
\citep[see, for example,][]{Donoho03,Chang06,Zhang07}, 
some of which are implemented 
in the publicly available source code for this work (see Appendix 
\ref{dimred_code}).

\section{An Application of LLE to Spectral Classification}
\label{LLE_on_spectra}

As a concrete example of the utility of LLE for astronomical 
classification, we apply LLE to spectra taken from the SDSS DR7 
data release \citep{SDSS-DR7}. 
A description of the SDSS can be found in \citet{York00} with the
underlying imaging survey described by \citet{Gunn98} and the photometric
system by \citet{Fukugita96}. The subsample used in this analysis was 
selected using the technique described in section \ref{sampling}.

This subsample comprises 8711 total spectra from DR7, with 6930
from the main galaxy sample \citep{Strauss02} and 1781 from the QSO sample 
\citep{Schneider08}, with redshifts z$<0.36$. All data
have been shifted to a common rest-frame, covering a spectral range of
3830 \AA\ to 9200 \AA, resampled to 1000 logarithmically-spaced
wavelength bins, corrected for significant sky absorption,
and normalized to a constant total flux. 
Emission-line and absorption-line equivalent 
widths have been extracted from the DR7 FITS headers. 
Using these data we further subdivide these spectra based
on the automated classifications of the SDSS spectroscopic
pipeline. Of the objects with Hydrogen emission line-strengths greater than
three times the spectral noise, 
888 objects are classified as broad emission line 
QSOs\footnote{Note: broad line QSOs in SDSS DR7 have \texttt{SPEC\_CLN=3}} 
(where broad means a line-width of $>1200$ km s$^{-1}$), 
893 are defined as narrow emission line QSOs, 
and 6068 are defined as emission-line galaxies (emission-line galaxies 
and narrow-line QSOs are distinguished using the [NII]/H$\alpha$
line-ratio diagnostic from \citet{Kewley01}).
Of the remaining galaxies, 523 are classified as absorption-line galaxies
(Balmer absorption strengths greater than 3$\sigma$), and 339 are classified as
quiescent galaxies (Balmer emission strengths less than 3$\sigma$).

\begin{figure}
  \centering
  \includegraphics[width=0.9\textwidth]{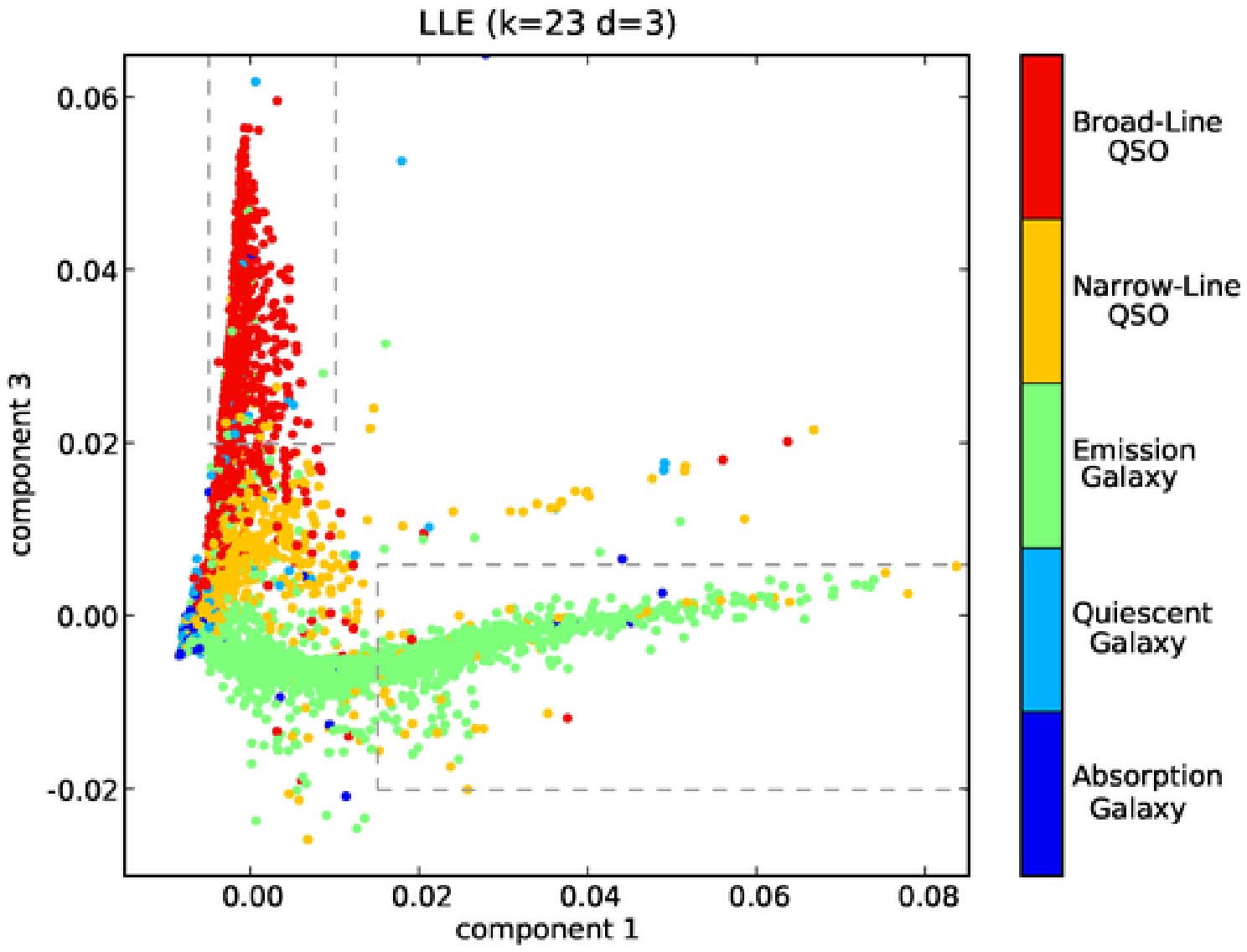}\\
  \includegraphics[width=0.45\textwidth]{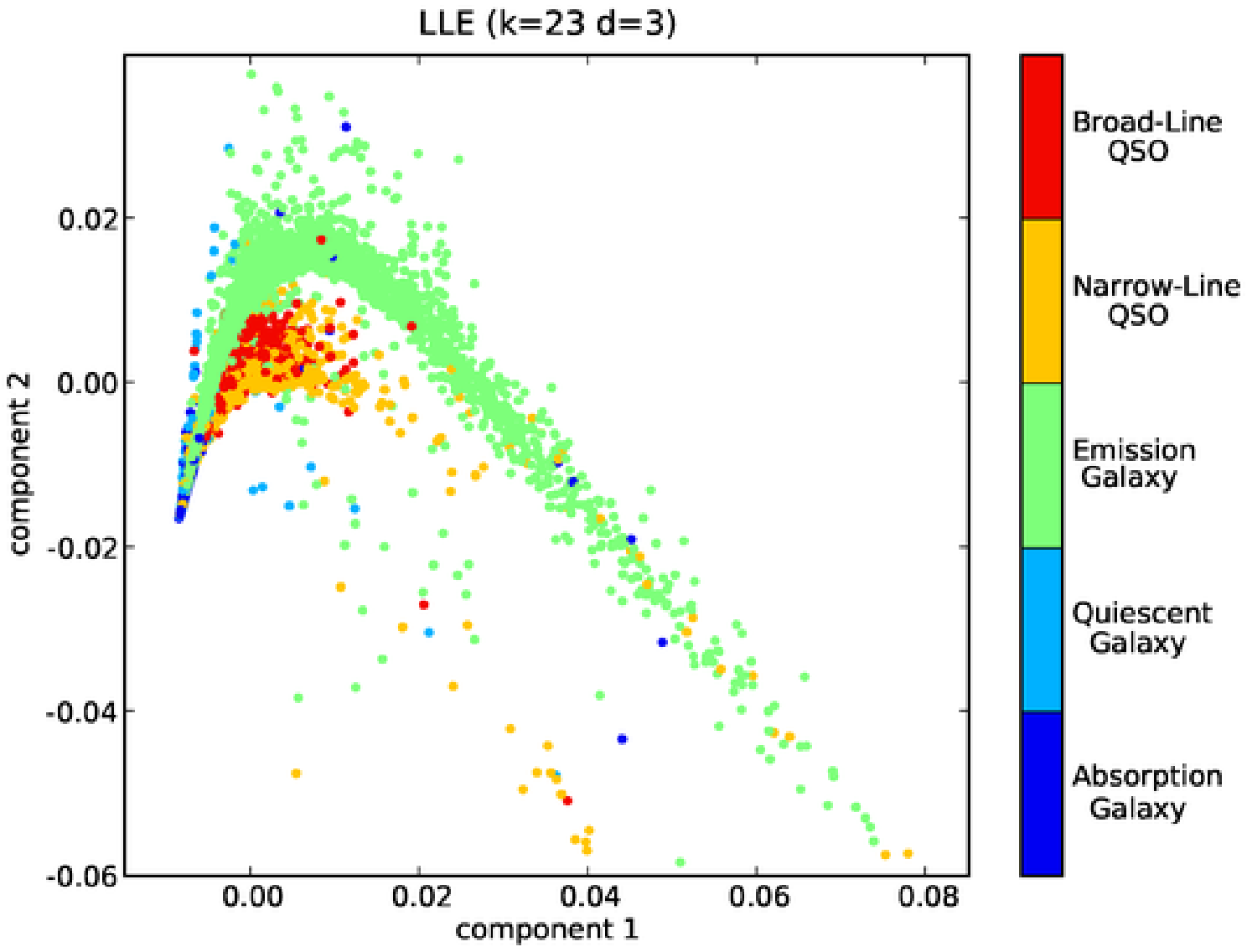}
  \includegraphics[width=0.45\textwidth]{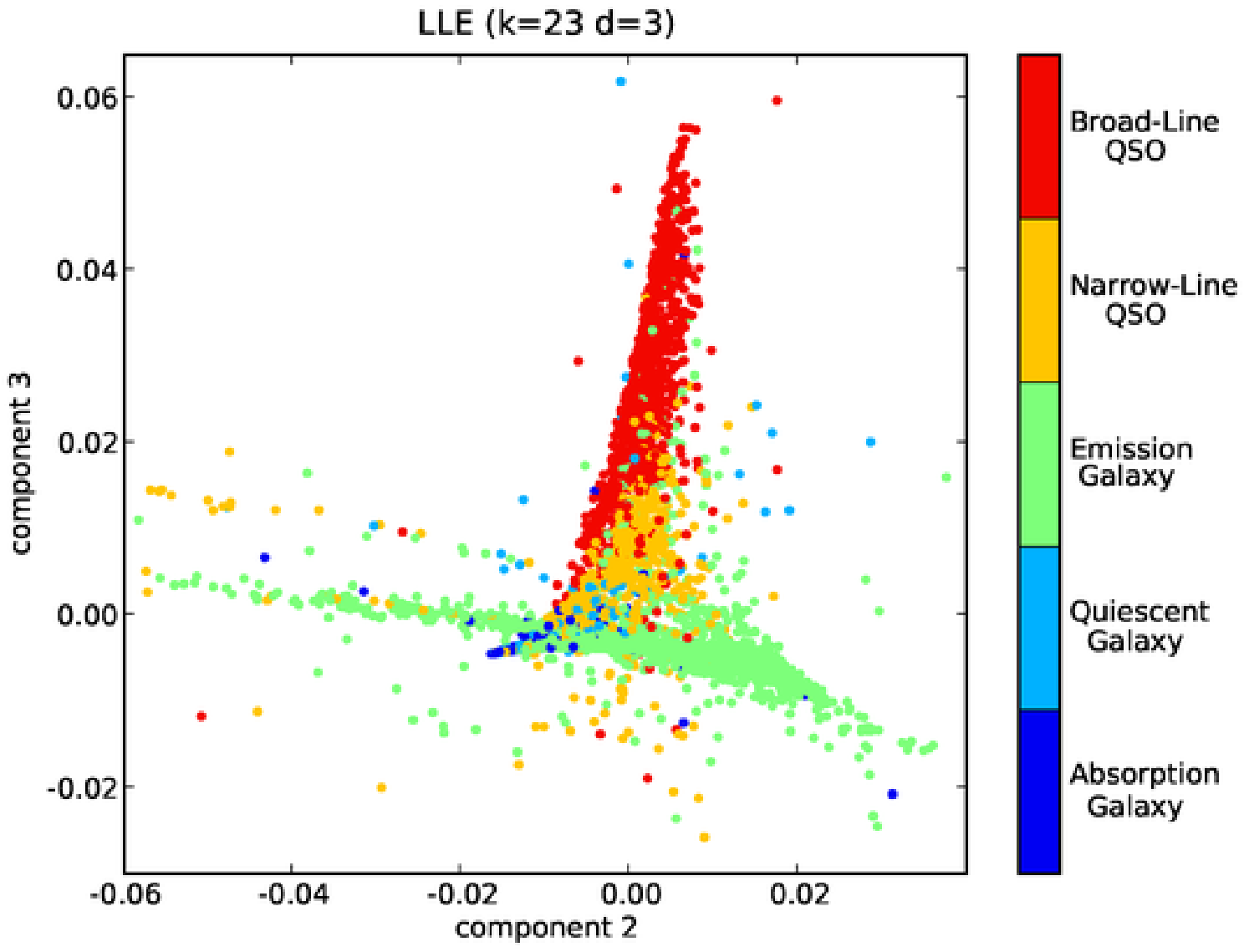} 
  \caption{Robust LLE projection of the data subsample from 1000 down to
    three dimensions.  
    Broad-line QSOs are identified by spectral classification within the
    SDSS pipeline, narrow-line QSOs and emission galaxies are distinguished
    using the [NII]/H$\alpha$ line-ratio diagnostic from \citet{Kewley01}, 
    and emission or absorption refers to galaxies with
    line strengths at three times the continuum noise.
    The dashed boxes indicate the ``broad-line QSO region'' and the 
    ``emission galaxy region'' discussed in section \ref{separation}
  }
  \label{LLEplot}
\end{figure}

Figure \ref{LLEplot} shows the LLE projection of the SDSS spectra from
the original 1000 dimensional space to a three dimensional
subspace. To minimize the effect of outliers, we use a robust variant of
LLE (see section \ref{RLLE}).  
We plot the position of all of the sources within this
subspace and color-code the points based on the above divisions. 
Red points are broad-line QSOs, orange points
represent narrow-line QSOs, green points are emission-line galaxies
and light and dark blue points are galaxies with continuum emission
and strong absorption line systems respectively.

The axes of an LLE projection are uniquely defined up to a scaling and
a global rotation. The interpretation of the projection comes,
therefore, from the relationship between the points rather than from
the axes themselves.  In Figure \ref{LLEplot}, it is clear that there
are a few well-defined branches in the resulting diagram that separate
the sources based on their spectral properties. QSOs (broad and narrow
line), emission-line galaxies and continuum galaxies separate in to
three basic clusters or classifications. Within these clusters broad
and narrow line QSOs can be distinguished as occupying separate parts
of the subspace, emission line galaxies are distinct from absorption
line systems while the strong absorption line systems closely track
the subspace of the continuum galaxies. The positions of galaxies with 
low emission and absorption line strengths converge to a common
region of the 3-dimensional space, which is to be expected as galaxies
do not represent a series of individual classifications but rather a
continuum of properties. 

For the remainder of the paper, all analysis will be done based on the 
two-dimensional manifold defined by the first and third components 
(i.e.\ the larger plot in Figure \ref{LLEplot}).  This sub-projection
gives maximal discrimination between the QSO and emission galaxy
branches (for further discussion of this choice of projection, 
see section \ref{choosing_parameters}).

\subsection{Interpretation of the LLE Tracks}
\begin{figure}
  \centering
  \includegraphics[width=0.8\textwidth]{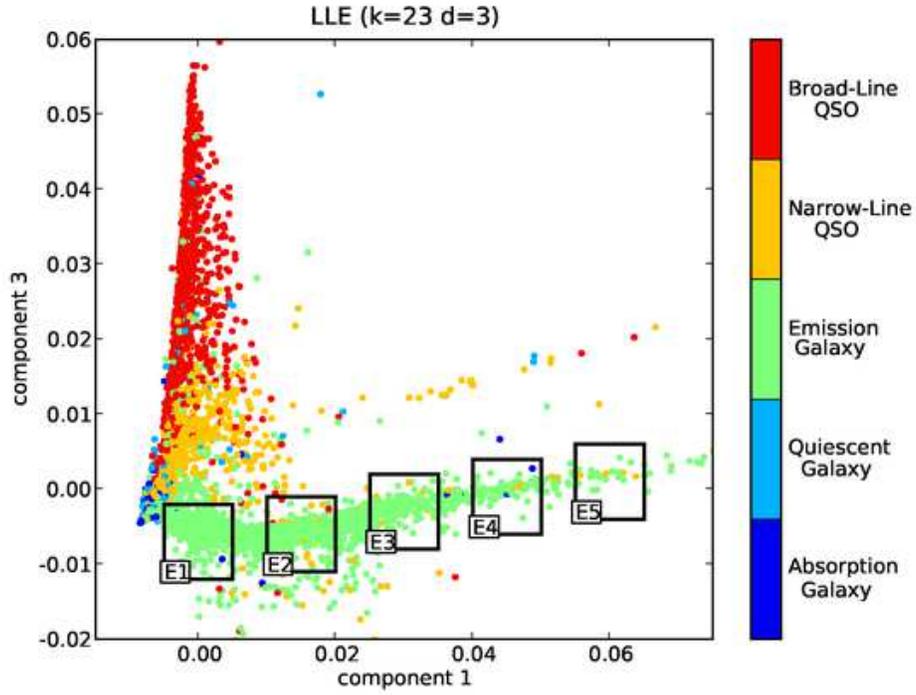}\\
  \includegraphics[width=0.8\textwidth]{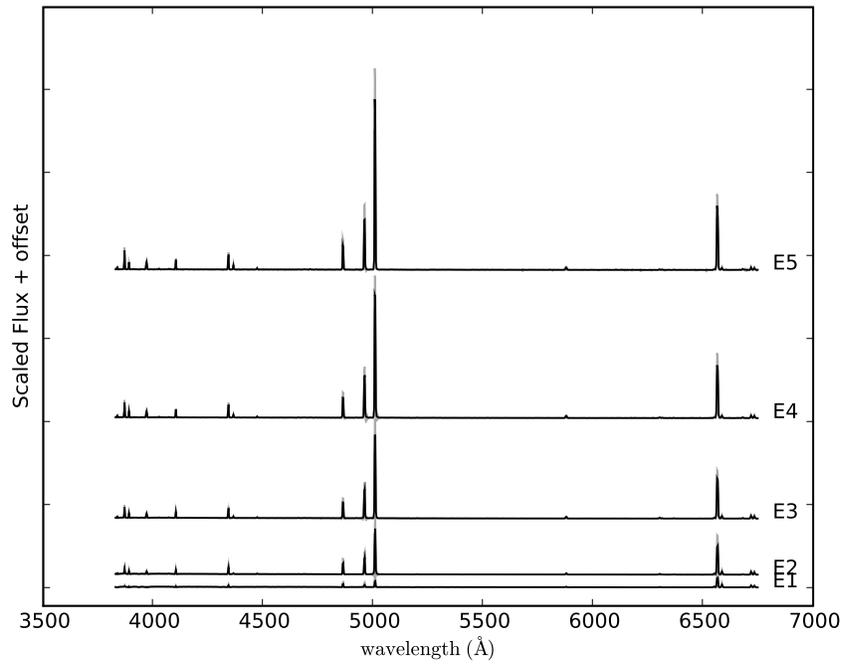}
  \caption{Progression of emission galaxy spectra.  Each spectrum represents
    the mean of the spectra within the labeled region.  Grey shading 
    indicates the standard deviation about the mean.
    See caption of Figure \ref{LLEplot} for description of colors.}
  \label{em_gal_track}
\end{figure}

\begin{figure}
  \centering
  \includegraphics[width=0.8\textwidth]{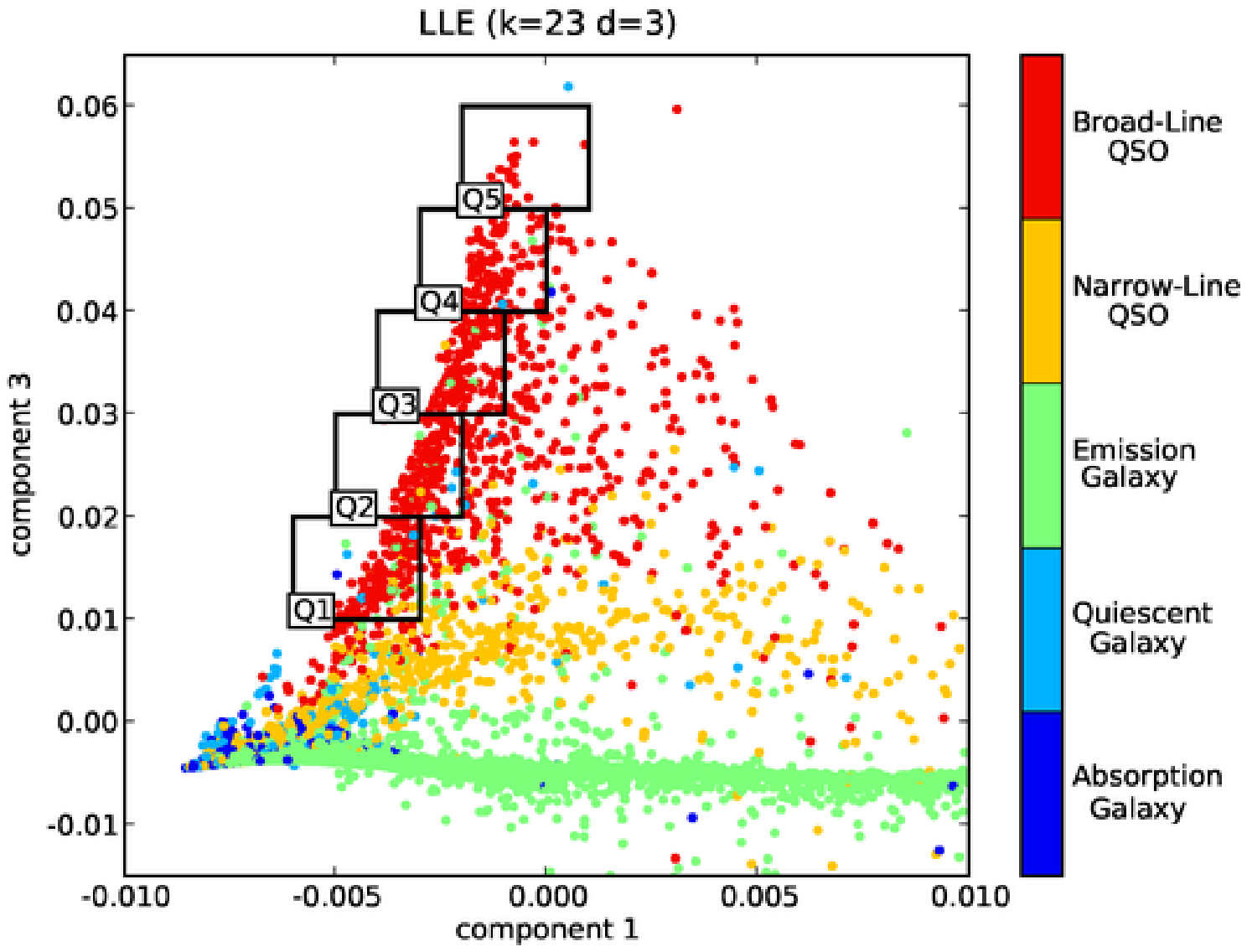}\\
  \includegraphics[width=0.8\textwidth]{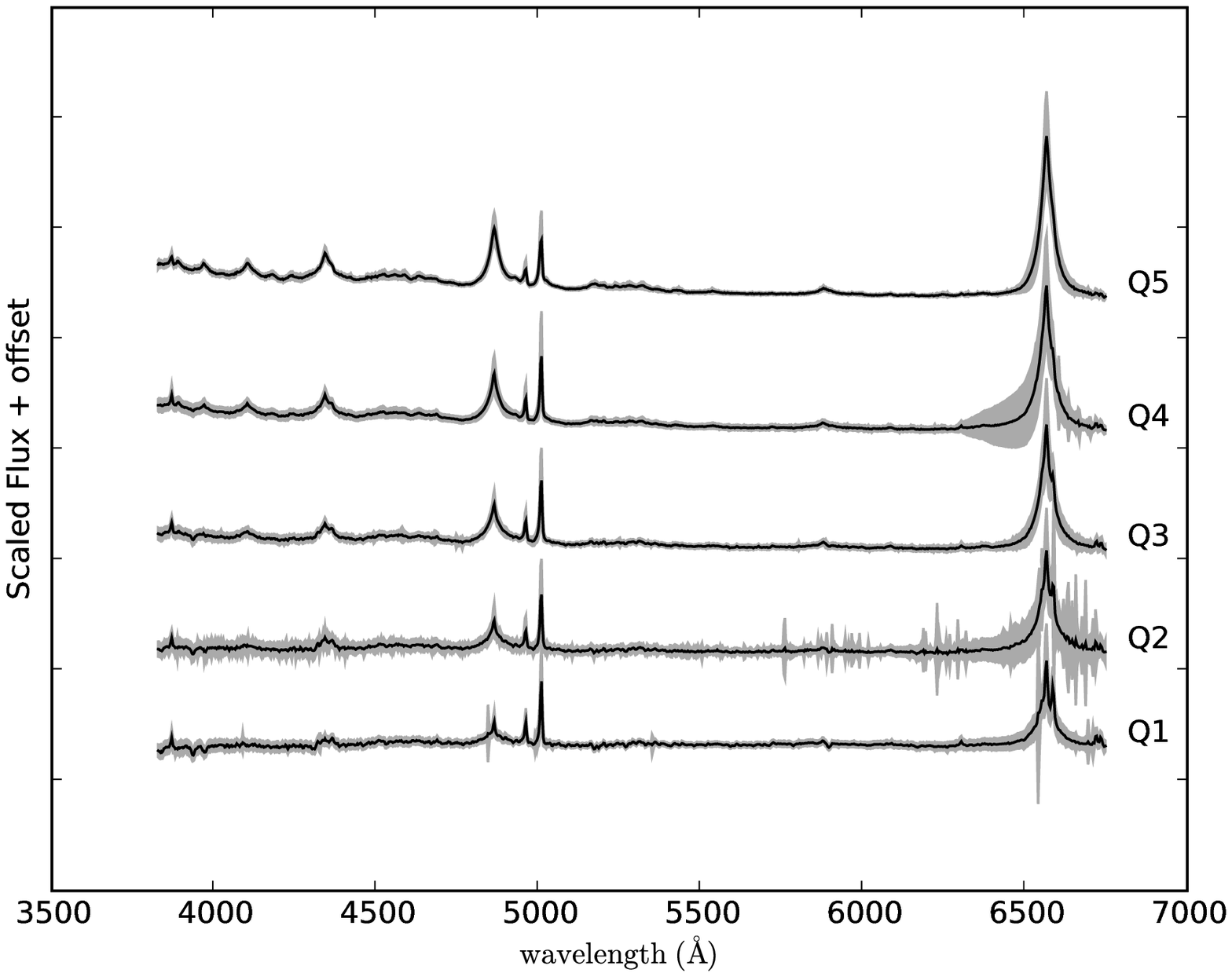}
  \caption{Progression of broad-line QSO spectra.  Each spectrum represents
    the mean of the spectra within the labeled region.  Grey shading 
    indicates the standard deviation about the mean.
    See caption of Figure \ref{LLEplot} for description of colors.}
  \label{broad_qso_track}
\end{figure}

\begin{figure}
  \centering
  \includegraphics[width=0.8\textwidth]{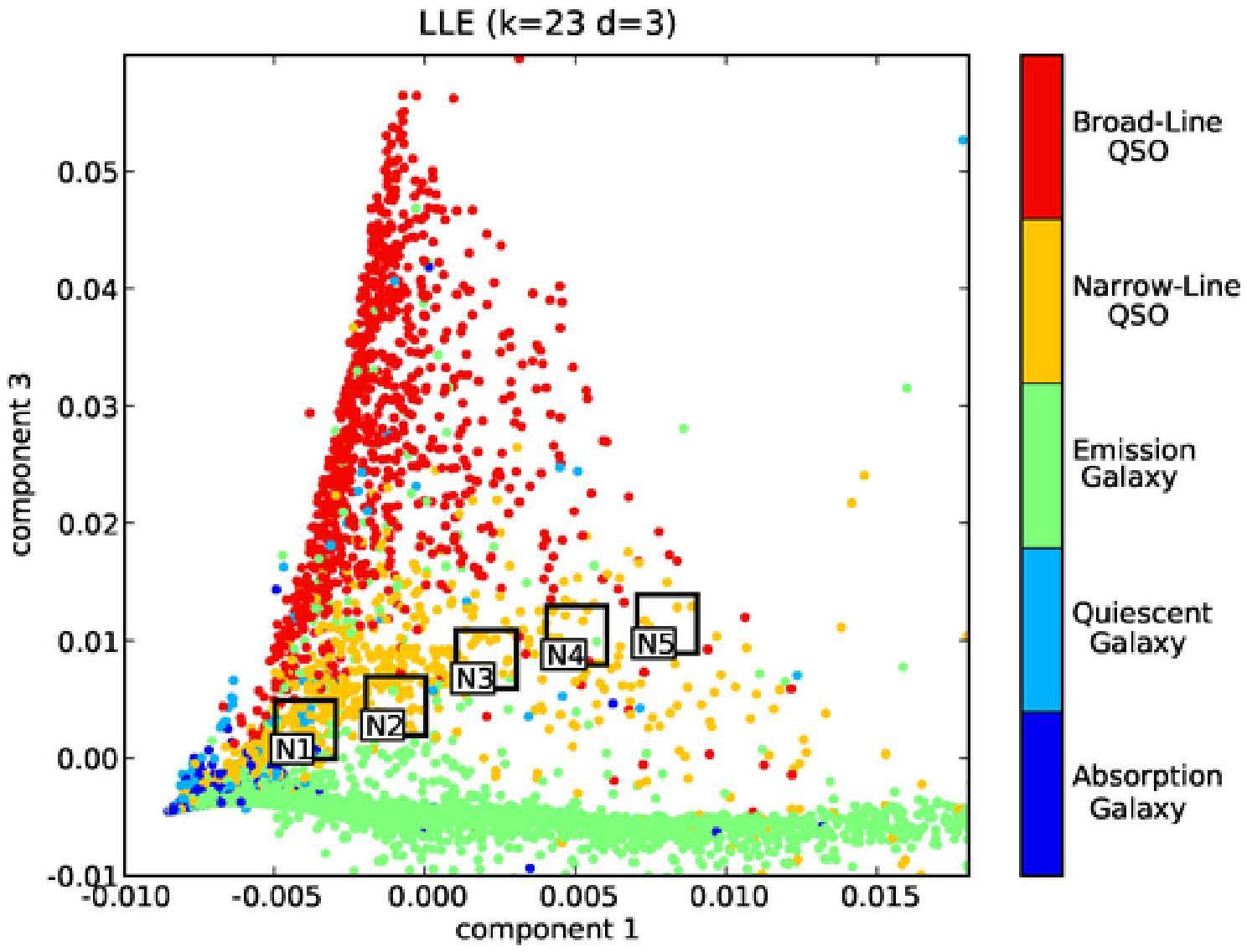}\\
  \includegraphics[width=0.8\textwidth]{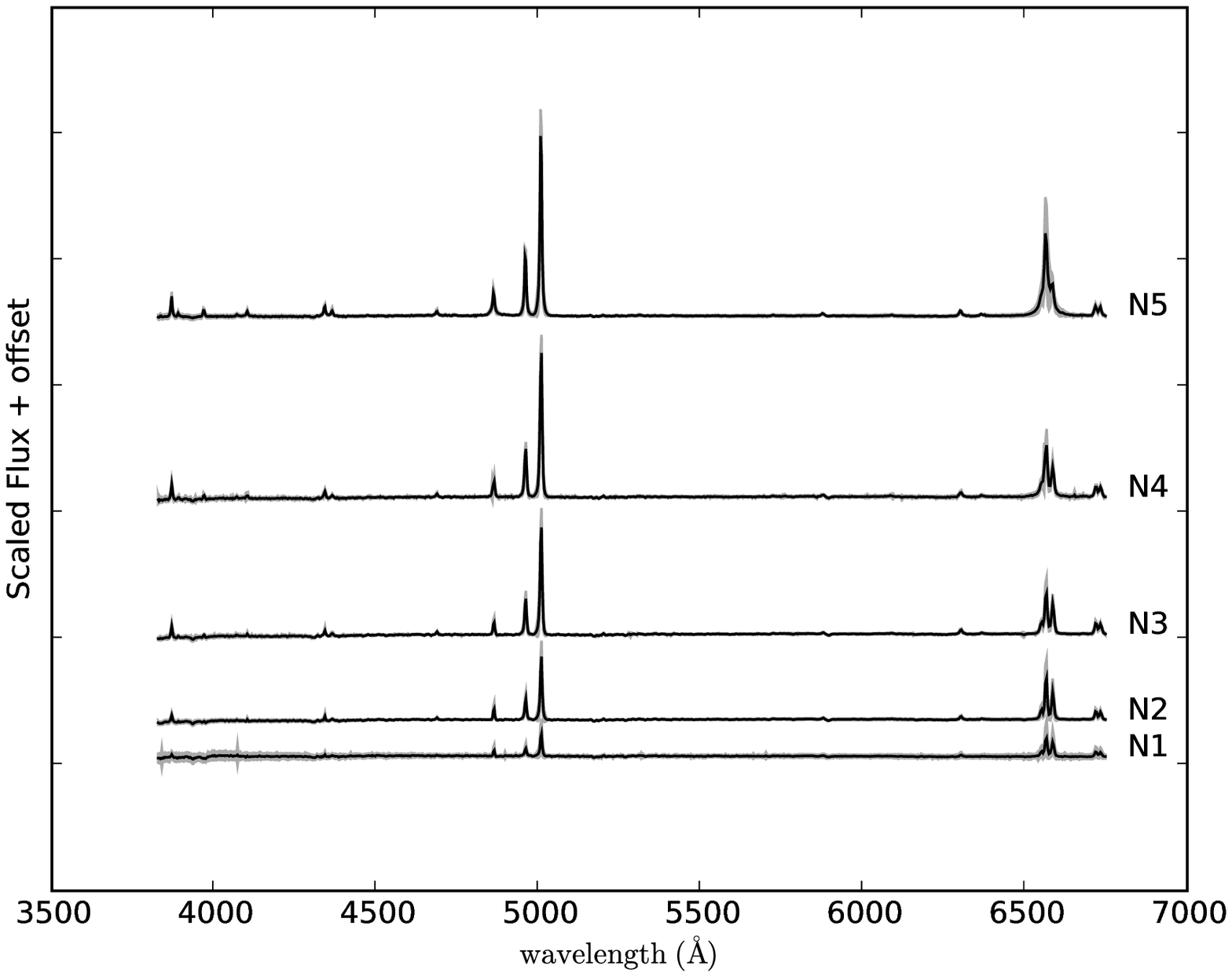}
  \caption{Progression of narrow-line QSO spectra.  Each spectrum represents
    the mean of the spectra within the labeled region.  Grey shading 
    indicates the standard deviation about the mean.
    See caption of Figure \ref{LLEplot} for description of colors.}
  \label{narrow_qso_track}
\end{figure}

\begin{figure}
  \centering
  \includegraphics[width=0.8\textwidth]{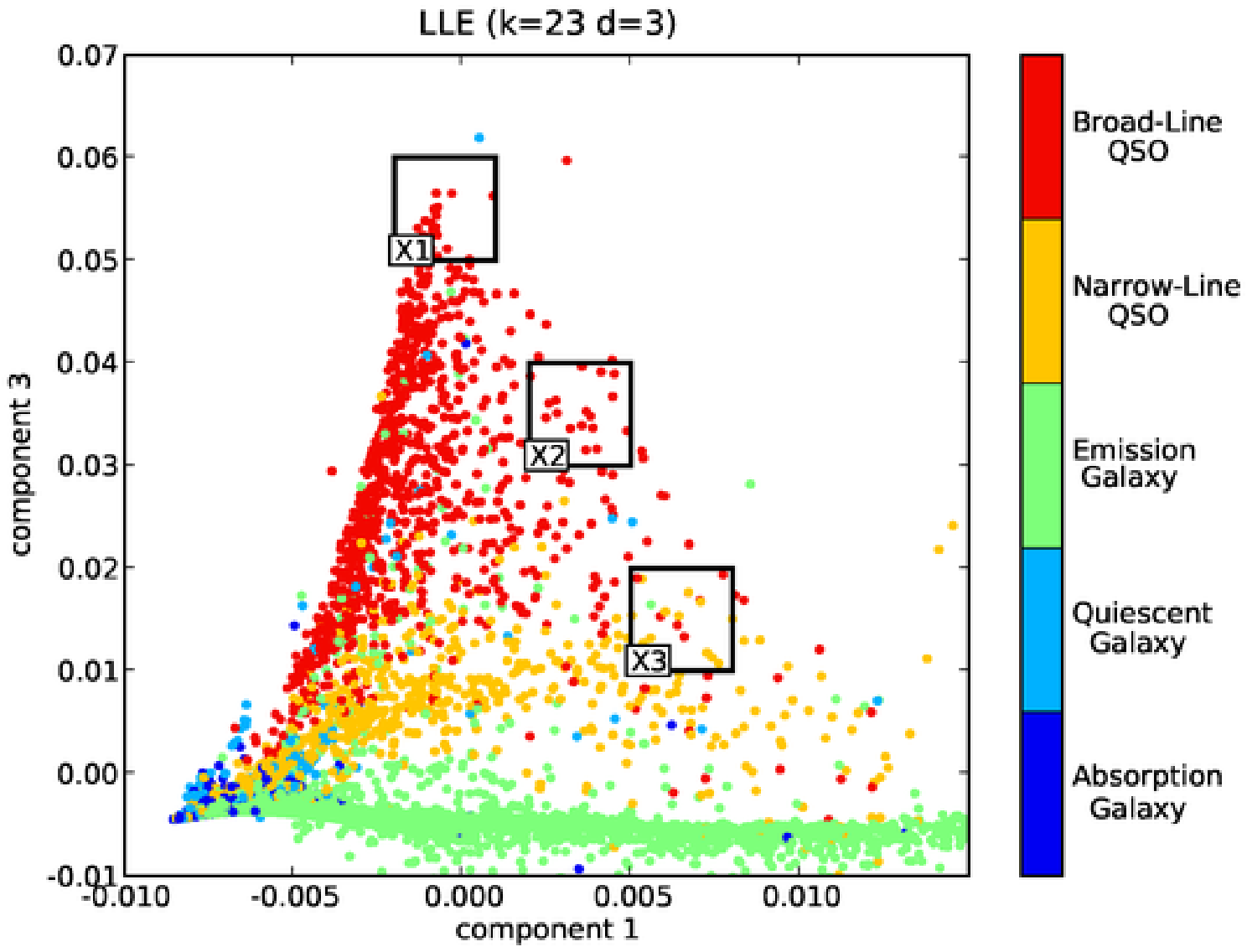}\\
  \includegraphics[width=0.8\textwidth]{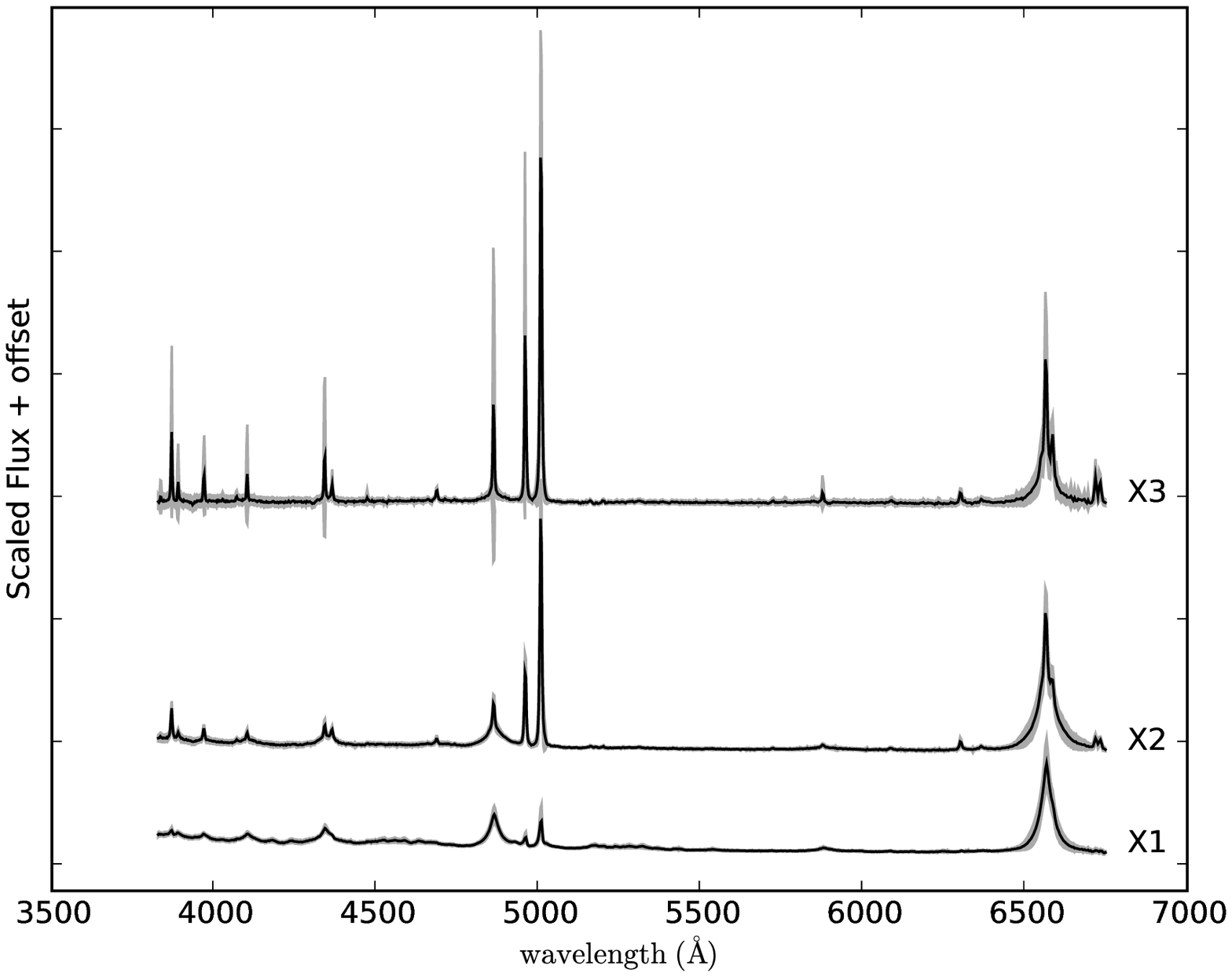}
  \caption{Progression from broad to narrow QSO spectra. Each spectrum 
    represents
    the mean of the spectra within the labeled region.  Grey shading 
    indicates the standard deviation about the mean.
    See caption of Figure \ref{LLEplot} for description of colors.}
  \label{qso_width_track}
\end{figure}

\begin{figure}
  \centering
  \includegraphics[width=0.8\textwidth]{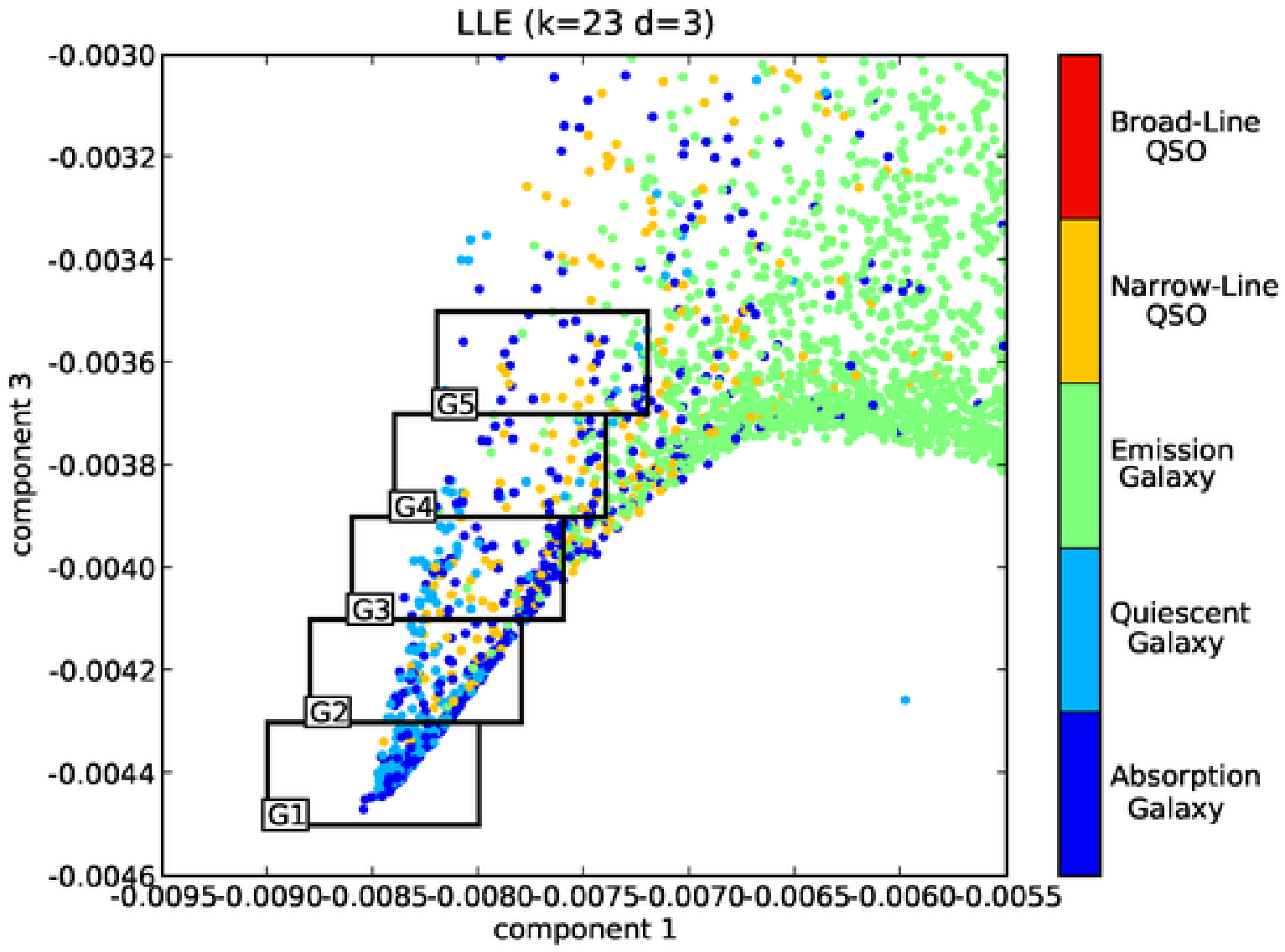}\\
  \includegraphics[width=0.8\textwidth]{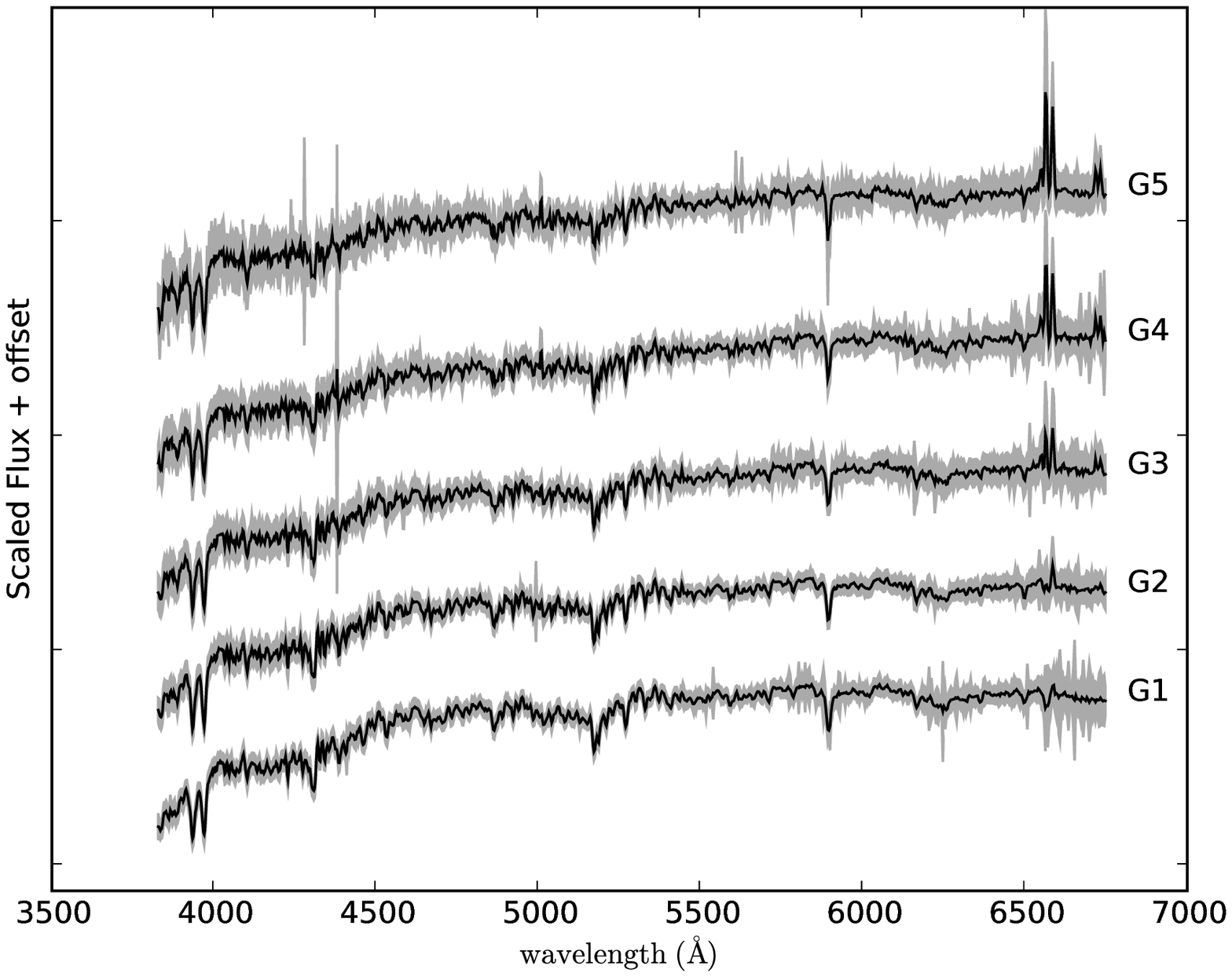}
  \caption{Progression of quiescent galaxy spectra. Each spectrum represents
    the mean of the spectra within the labeled region.  Grey shading 
    indicates the standard deviation about the mean.
    See caption of Figure \ref{LLEplot} for description of colors.}
  \label{quiescent_gal_track}
\end{figure}

The distance of a galaxy along any one of the branches within Figure
\ref{LLEplot} correlates well with its physical characteristics (as
reflected in their spectra).  Figures \ref{em_gal_track} through 
\ref{quiescent_gal_track} show progressions along various regions, along
with the mean spectrum of each labeled region. When calculating the mean,
we exclude excessively noisy objects, identified as those with flux
$F(\lambda)$ satisfying 
$\int|F(\lambda)|d\lambda > 1.5\int F(\lambda)d\lambda$. 
Because the spectra are preprocessed with a normalization, the few
spectra satisfying this inequality would otherwise contribute 
disproportionately to the mean spectrum in each region.

Figure \ref{em_gal_track} shows the
progression of spectra along the emission galaxy track. To do this we
calculate the mean spectrum for galaxies contained within the
regions associated with the bounding boxes E1 through E5. As we will show
later, this sequence not only tracks increasing line strengths, but also 
changing line ratios (see Figure \ref{BPTplot}). 

Figure \ref{broad_qso_track} shows the progression of spectra along
the broad-line QSO track.  We find that, as we move from Q1 to Q5, the
spectral type of the QSO evolves from a Seyfert 1.9 to a classic broad
line QSO. As would be expected for such a transition the NII emission line
strength decreases while the H$\beta$ line strength increases as the
spectra become progressively dominated by the accretion disk
emission. Associated with these emission line properties we find that
the continuum slope of the QSO spectra are bluer for the Q5 class than
for Q1. This not only supports the classification of the spectra from
Seyfert 1.9 to broad-line AGN, it also demonstrates how LLE can use
jointly the information contained within the continuum and line
emission.

Figure \ref{narrow_qso_track} shows the progression of spectra along the
narrow-line QSO track.  As with the emission-line galaxy track, the
narrow QSO track traces increasing emission from N1 to N5.  N1-N3 have narrow
H$\beta$ and H$\alpha$, with line-ratios that indicate power-law ionization,
and are spectrally consistent with LINERs.  N3-N5 have increasingly broad
H$\alpha$, which suggests classification as Seyfert 1.9-2.0.  Line-ratio 
diagnostics confirm this (see Figure \ref{BPTplot}).  


In Figure \ref{qso_width_track} we consider the progression of spectra
across the different spectral groupings (from broad line QSO to
galaxies with narrow emission lines) as opposed to along an individual
track. The X3 sources are, as before, broad-line QSOs with strong
H$\beta$ emission. As we transition to X2 the width of the emission
line decreases until X2 has the spectral characteristics of a Seyfert
type 1.5. By X1 the spectral properties are consistent with an HII
region or narrow emission line galaxy (for wavelengths $<$ 5500 \AA)
while showing evidence for an AGN component in the broad line
H$\alpha$. This one parameter sequence from broad-line QSO to HII
region spectra (including regions where the populations are mixed) is
described very naturally with this LLE projection.

The last subsample we consider are the quiescent galaxies in Figure
\ref{quiescent_gal_track}. The transition from G1 to G5 tracks the
same spectral classification from quiescent to active star formation
as found using the PCA analysis of \citet{Yip04}. In terms of
spectral properties, galaxies classified as G1 correspond to the E/S0
spectral type from \citet{Kennicutt98} and galaxies in 
G5 correspond to the Sbc/Sc
Kennicutt spectrum. Along this progression the emission line strengths
for H$\alpha$ and NII increase and the strength of the MgIII
absorption line decreases. It is also noted that the depth of the
break at 4000 \AA\ decreases with increasing index number. Comparisons
with the results of \citet{Yip04} indicate that this trajectory maps
directly to the star formation rate within the galaxy (see Figure 
\ref{PCA_plot}).

From these initial comparisons it is clear that by projecting onto a
three dimensional space using LLE we can differentiate between 
line-emitting galaxies and those dominated by continuum emission. We can also 
distinguish between sources as a function of their emission line 
characteristics. As such, LLE encapsulates both the line emission and 
continuum emission properties in its classification scheme. That LLE can 
express the variation in continuum properties of galaxies when projecting 
from 1000 to 3 dimensions is not altogether surprising as the PCA analyses
of \citet{Connolly95}, \citet{Yip04} and others have demonstrated that
quiescent galaxies occupy only a small region of the available parameter space
and that the dominant direction in this space is governed by the star
formation rate. For emission line galaxies, and in particular QSOs,
standard PCA approaches require between 10 and 50 components to
express the variation in line strengths as a function of galaxy
type. This is due to the inherently non-linear nature of the
reconstruction of emission lines. 

The fact that LLE can be used to separate both
broad and narrow emission lines into separate groups or families, that
the transition between narrow and broad line spectra is smooth, and
that the position of a galaxy within each of these groups is dependent
on the individual line strengths demonstrates the ability of LLE 
to concisely represent inherently non-linear systems in an intuitive manner.

\subsection{Separation of Spectral Types} 

\label{separation}

LLE is useful in that it maps a large-dimensional dataset to a smaller
dimensional parameter space.  Within this projected space, many familiar
classification schemes can be applied \citep[see][for a thorough introduction
to the topic]{Hastie01}.  In the case of SDSS spectra, LLE succeeds in mapping
physically similar spectra to the same region
of parameter space.  More importantly, physically different spectra are
mapped to distinct, well-separated regions, leading to the possibility
of a very intuitive classification scheme.

\begin{figure}
  \centering
  \includegraphics[width=0.8\textwidth]{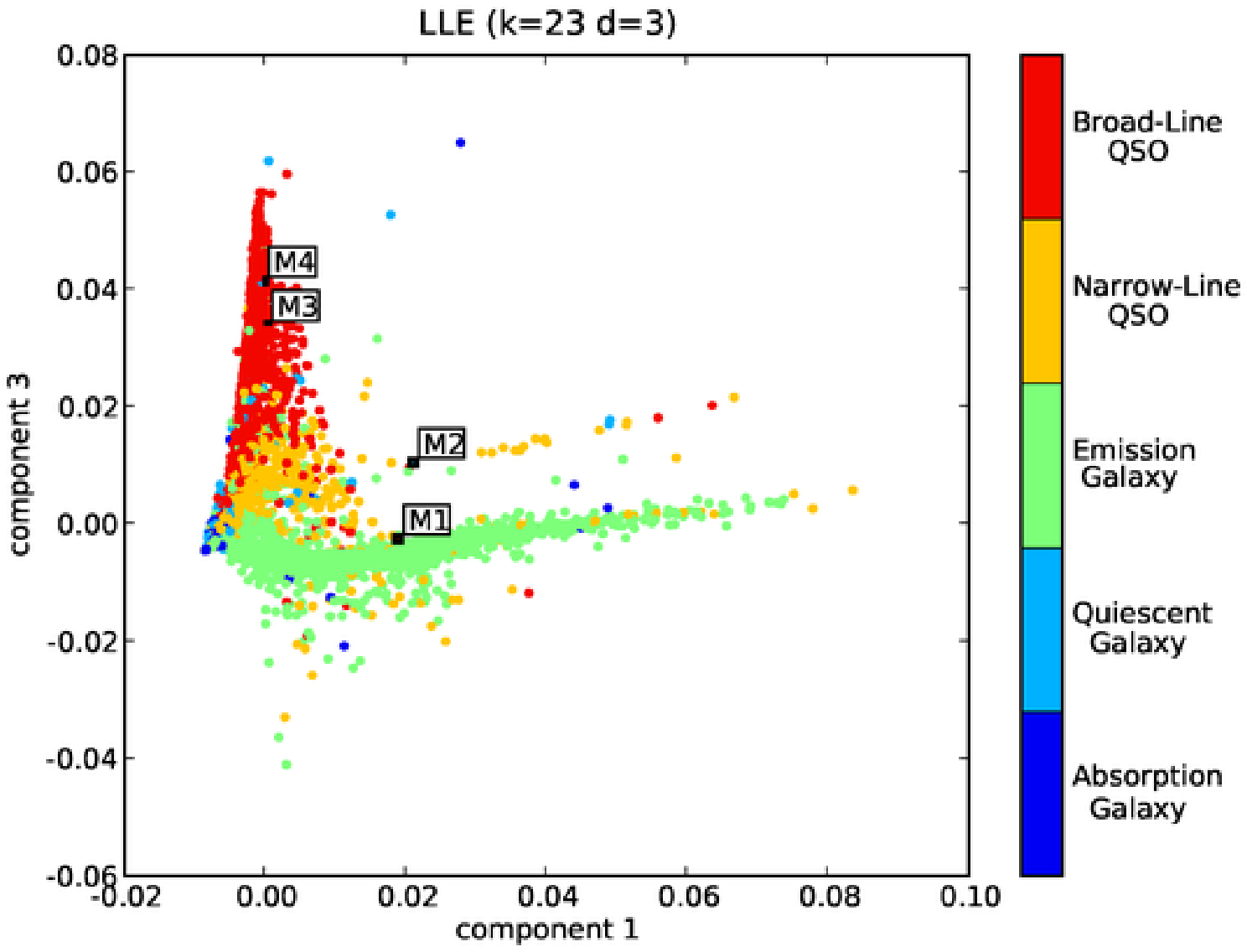}\\
  \includegraphics[width=0.8\textwidth]{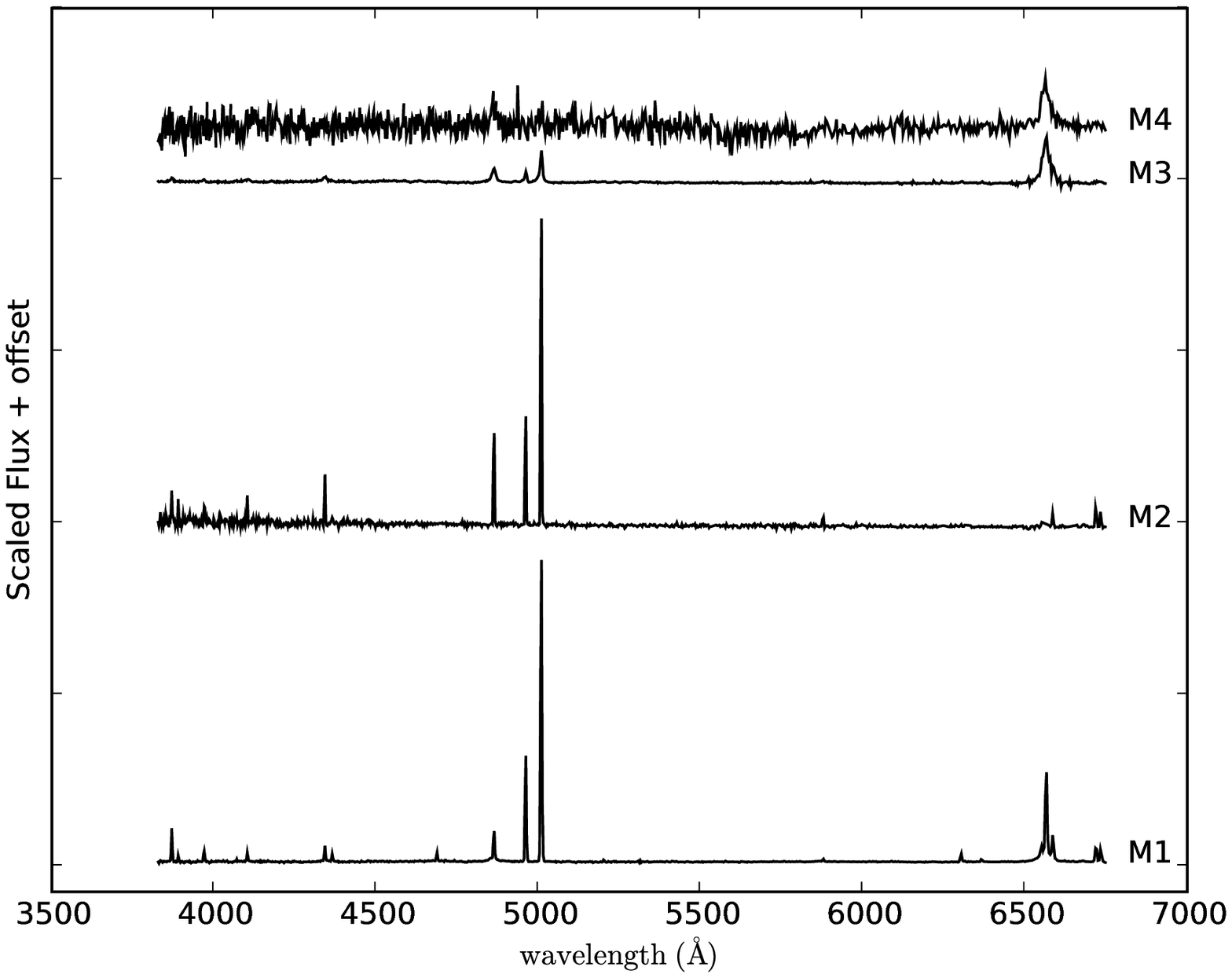}
  \caption{A sample of objects with discrepant classifications
    See caption of Figure \ref{LLEplot} for description of colors.}
  \label{misclassified}
\end{figure}

Because of this, we can quickly locate objects that are
misclassified by both the SDSS pipeline and the line-diagnostic plots. 
Figure \ref{misclassified} displays a few examples of these.  M1 is 
classified by the SDSS pipeline as a broad-line QSO, but falls on the
emission-line galaxy track.  This mis-classification is likely due to
contamination of the H$\alpha$ line from the neighboring Nitrogen features,
leading to an erroneous large line-width.  M2 is not recognized as an emitting
object, because the H$\alpha$ feature is missing.  In fact, the entire 
narrow track on which M2 lies consists of objects which look like emission
galaxies/type-2 Seyferts, with the H$\alpha$ line missing due to 
saturation of the lines, coincident sky lines, and bad pixels.  
LLE isolates these points on their own track of outliers.
M3 and M4 fall on the broad-line
QSO track, but are not recognized as broad emitters by the SDSS pipeline, 
probably due to high background noise. The LLE projection correctly groups
these with other strong emitters.  Because the LLE projection is entirely 
neighborhood-based, any given point will have similar properties to nearby 
points in the projection space. 

Next we compare the classification based on LLE to the classification by more
traditional methods, e.g.\ emission line-widths, line-strengths, and 
line-ratios. This requires a choice of classification 
technique for points in the LLE-projection space.  
For illustrative purposes, we create a simple classification scheme and
define two well-separated
regions which correspond to physically different objects.  The
``broad-line QSO region'' is defined as all points on the broad-line track 
with LLE component three greater than 0.02.  The ``emission galaxy region'' is
defined as all points on the emission galaxy track with LLE component one
greater than 0.015 (see figure \ref{LLEplot}).

Out of the 624 galaxies in the broad-line QSO region, 31 were classified by
SDSS as having narrow line emission, and 11 were classified as quiescent.
Through visual inspection, 29 of the 31 and 1 of the 11 were found to show 
broad emission features, with the rest of the spectra too noisy to 
accurately classify.  The SDSS pipeline, therefore, mis-classifies about
5\% of their broad-line sources as narrow-line.

Out of the 675 galaxies in the emission galaxy region, 45 have 
line ratios on the narrow-line QSO side of the \citet{Kewley01}
cutoff, two have H$\alpha$ emission broader than 1200km s$^{-1}$, and 
4 have prominent emission features not detected by the SDSS pipeline.

Combining the above tallies, we can estimate how successfully LLE clusters
physically similar spectra.  From a subsample which by design contains
a high fraction of abnormal spectra (see Section \ref{sampling}), less than
one percent of the spectra were physically dissimilar from 
the average population in each region.
These errors are due primarily to very noisy spectra. 
For other cases, e.g.\ the 45 spectra
where classifications based on LLE and \citet{Kewley01} line diagnostics 
disagree, it is not immediately clear which classification is correct, if any.
The majority of these sources are transitional,
falling near the dividing boundary of the line-ratio
diagram.  This dividing line is an arbitrary analytic separation 
between two regions of a certain subset of the parameter space.  
We note, however, that the LLE-based classification is without any 
\textit{a priori} training.  We do not preselect sources based on
known properties nor fine-tune the parametrization of the classification,
as opposed to the optimized SDSS pipelines.
 
\begin{figure}
  \centering
  \includegraphics[width=0.8\textwidth]{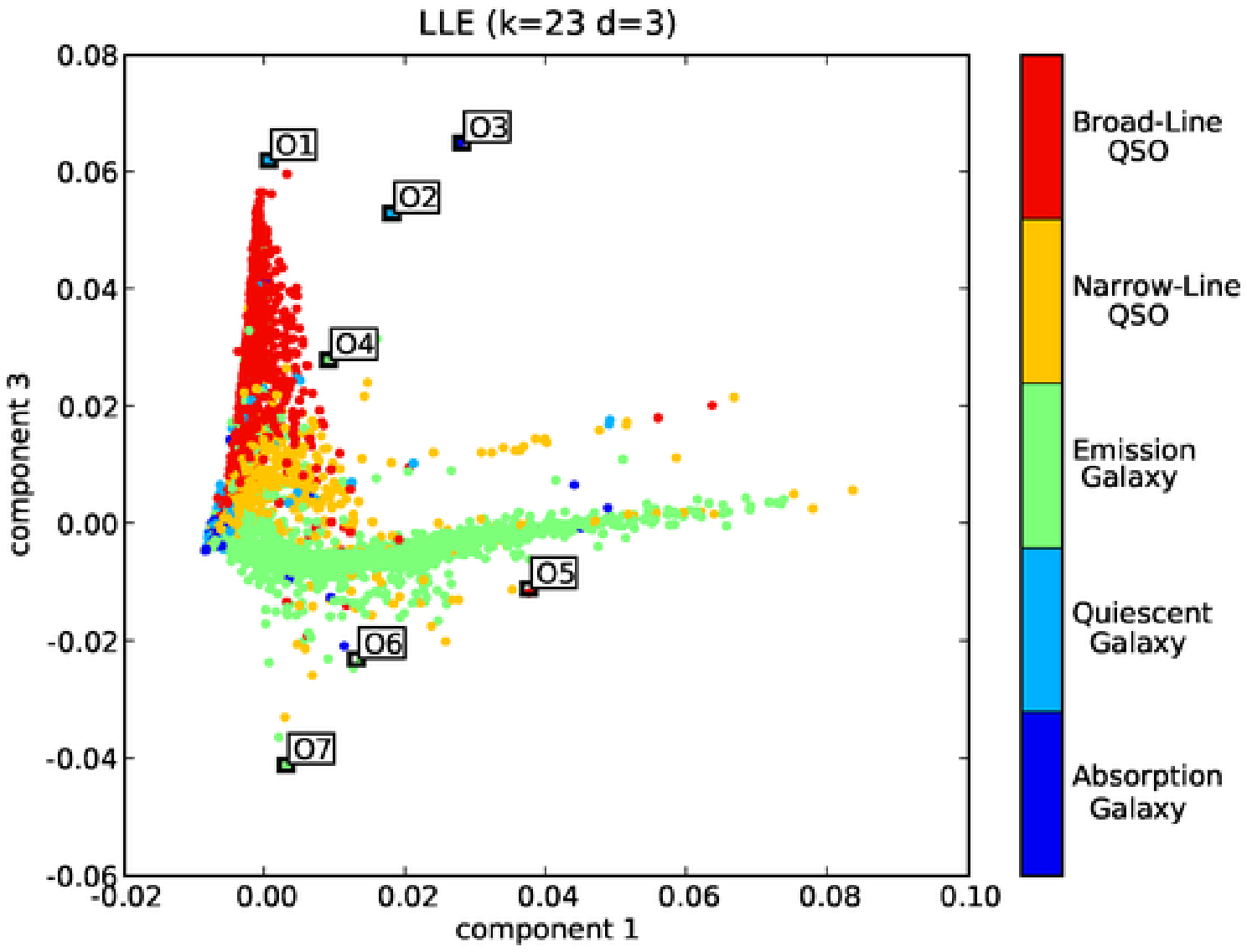}\\
  \includegraphics[width=0.8\textwidth]{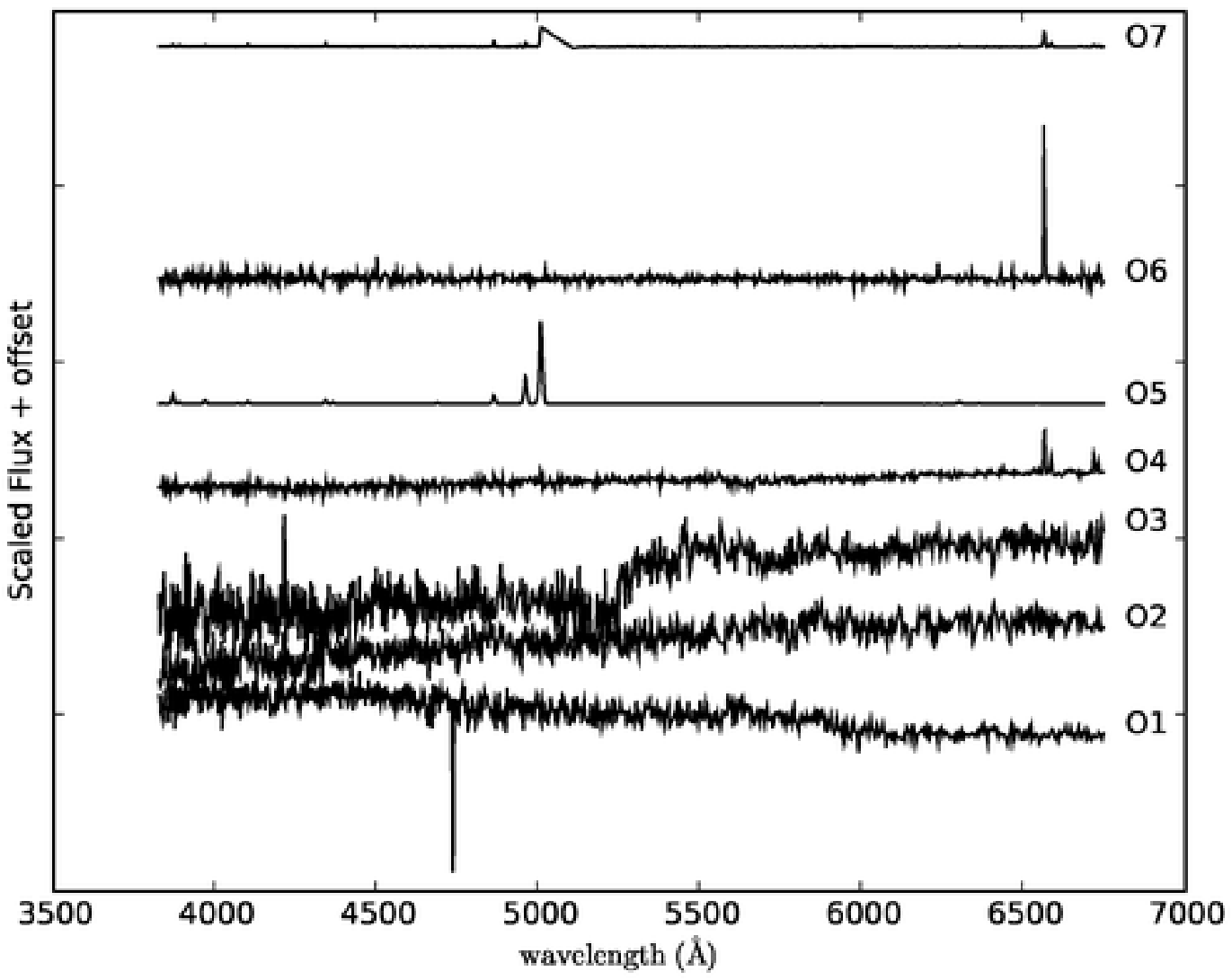}
  \caption{A sample of outliers from the LLE projection
    See caption of Figure \ref{LLEplot} for description of colors.}
  \label{outliers}
\end{figure}

Figure \ref{outliers} shows a sample of some significant outliers
identified with LLE.  O1-O3 are outliers due to their noise.  O4 and O6
appear to be sky spectra which were misclassified based on 
emission features.  O5 appears to be a QSO with its H$\alpha$ line
missing, possibly due to atmospheric absorption.  O7 has an artificial
feature likely due to faulty sky-noise subtraction.  In all of these cases, the
spectrum's status as an outlier based on LLE reflects its unusual 
spectral characteristics.


\section{Considerations when applying LLE}

\label{LLE_considerations}

\subsection{Robust LLE: dealing with noisy and missing data}
\label{RLLE}
As with any classification scheme it is important that we can account
for missing and noisy data in a robust manner. As LLE is a locally
weighted projection it is expected that outliers (even outliers for a
single pixel) can impact the definition of the local neighborhood and
might bias any classification scheme.

In practice it is difficult to treat this problem completely as it
essentially replaces the least squares analysis of each neighborhood
with a \textit{Total Least Squares}/\textit{Errors in Variables}
 analysis for each of
the $N$ points.  These cannot be solved in closed-form, and require a
computationally expensive iterative process \citep[see][for a detailed
exploration of the Total Least Squares problem]{TLS}.

We can, however, follow the methodology of \citet{Chang06} and address the
problem of outliers using a Robust LLE (RLLE) technique.  In RLLE, a
\textit{reliability score} for each data point is determined based on its
efficacy in predicting the location of the points in its neighborhood.
Outliers will have two general properties: first, they will not be members
of large number of local neighborhoods; second, within the neighborhoods
in which they appear, they will be far from the best-fit hyperplane.
With this in mind, reliability scores are calculated using the following 
method:
\begin{enumerate}
\item For each point, an \textit{iteratively reweighted least squares}
  analysis \citep{Holland77} is performed to determine optimal weights 
  for a weighted PCA reconstruction of the point from its neighbors. 
  Schematically, this is akin to finding the best-fit hyperplane,
  weighting points based on their distance from that hyperplane, and
  repeating until the process converges.  As a result of this process,
  each point in a local neighborhood is assigned a local weight which
  quantifies its contribution to the local tangent space.
\item Each point may be a member of many neighborhoods. The reliability
  score for a point is determined by taking the sum of its local weights 
  from each neighborhood of which it is a member.
\end{enumerate}
Thus, a higher score will be given to points which are part of many 
neighborhoods, and points which contribute more to the reconstruction
in each neighborhood.  With this in mind, outliers can be identified 
based on their low reliability score.

\begin{figure}
 \centering
 \label{RLLE-ex}
 \includegraphics[width=\textwidth]{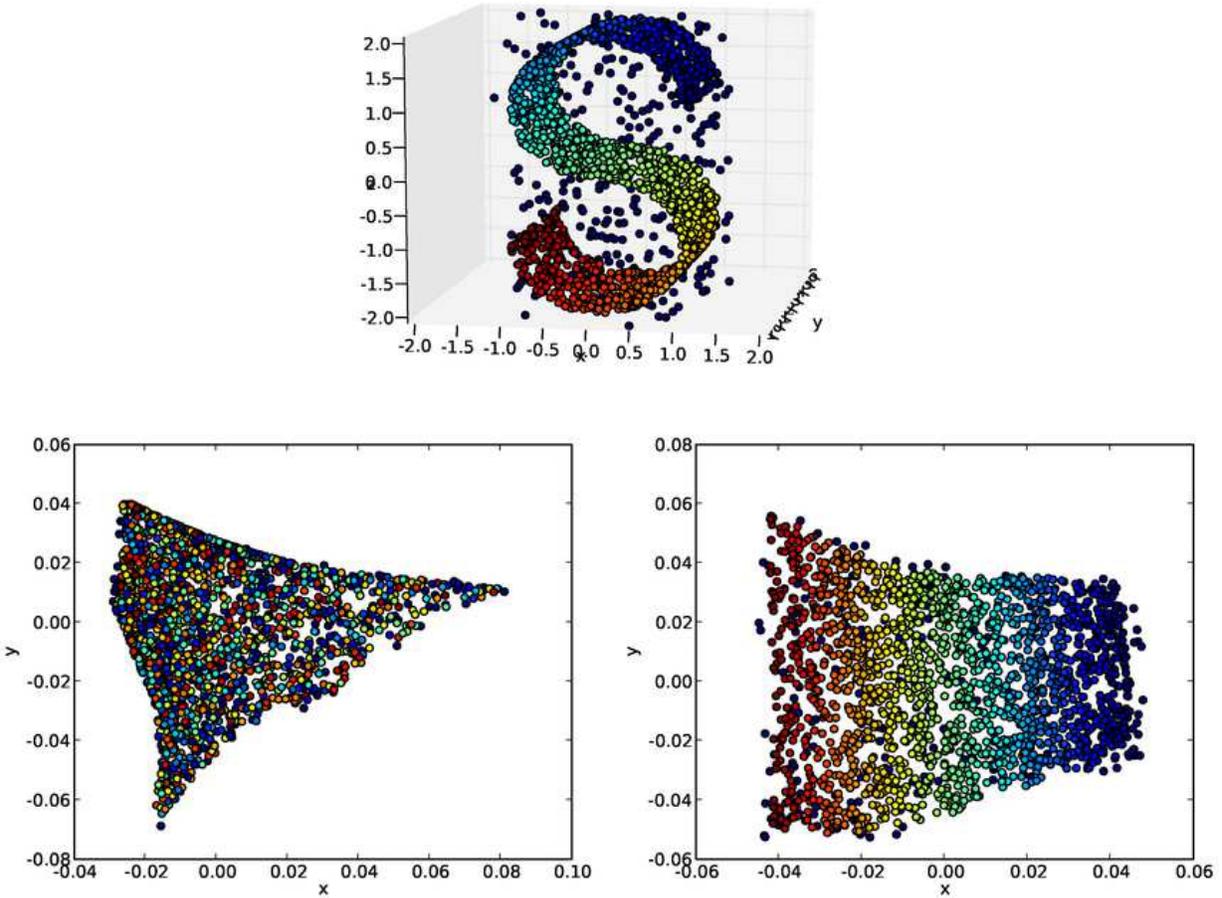}
 \caption{\textit{Top:} $N=2000$ points randomly sampled from a bounded 
   2-manifold embedded in 3-space, with 15\% of the points random 
   outliers sampled from the entire parameter space.  
   \textit{Bottom Left:} the 2-dimensional LLE projection of this manifold, 
   with $k=15$.  Note the mixing of the colors: the outliers obscure the
   2-dimensional manifold to the point that LLE cannot recover it.
   \textit{Bottom Right:} the 2 dimensional RLLE projection of this manifold, 
   with $k=15$, constructed using 80\% of the points with the highest
   reliability scores.  RLLE recovers the desired projection.}
\end{figure}

In Figure \ref{RLLE-ex} we compare the performance of LLE
and RLLE on the S-curve dataset from section \ref{simple-example}. 
In this case we assign 15\% of the points within the data
set to be randomly selected from the full parameter space. The lower
left panel shows the application of the standard LLE approach. As is
clear from this Figure, outliers can change the local weights of the
nearest neighbors and, thereby, distort the resulting lower
dimensional manifold; causing the points of different colors to be
heavily mixed. Effectively, the random noise within the point
distribution fills in enough of the three-dimensional space that
LLE cannot extract the lower dimensional manifold.  For RLLE, as long
as the number of random points is small relative to the number of
nearest neighbors, we can isolate and exclude the contaminating points
from the local weights. Figure \ref{RLLE-ex} shows that we can recover
the underlying two-dimensional manifold in the presence of noise and
outliers through this approach.

\subsection{Choosing Optimal Parameters}
\label{choosing_parameters}
There are three free parameters that must be chosen for LLE: $K$, the number
of nearest neighbors, $D_{out}$, the output dimension, and $r$, the 
regularization parameter.  $K$ and $D_{out}$ will be discussed here, while
$r$, which depends on details of the implementation, is addressed in 
appendix \ref{finding_w}.

Often in PCA, instead of specifying an output dimension $D_{out}$, one
chooses to specify a percentage of the sample variance which should be
preserved.  In LLE, dimensionality can be determined in an analogous
way for each local neighborhood \citep{deRidder2002}.  
In the computation of an LLE
projection, a variant of the covariance matrix is computed for each
local neighborhood.  This matrix differs from the covariance matrix
used in PCA by only a translation and scaling, neither of which affect
the relative magnitudes of the eigenvalues.  Taking the lead from PCA,
we can specify a percentage of the \textit{local} variance which we
would like to conserve, and thereby find the minimum number of
dimensions required locally, simply by doing an eigenanalysis of the
local covariance matrix.  Taking the mean of the $N$ local
dimensionality determinations gives an estimate of the optimal value
of $D_{out}$ for the global projection. For our SDSS analysis, the first three
components contain an average of 75\% of the local variance.  The first
ten components contain 92\% of the local variance, significantly less than
the 99.9\% of total variance within the first 
ten PCA eigenspectra of \citet{Yip04}.
This discrepancy is likely due to noise and intrinsic variation
at the local level, which contributes more to the small, local 
neighborhoods which LLE probes than to the overall variance that PCA probes.
Also, our sampling technique (Section \ref{sampling}) chooses sources
based on dissimilarity, leading to more intrinsic variation in the 
subsample.

Choosing the number of nearest neighbors is more difficult.  In fact,
$K$ need not be constant across all neighborhoods.  It could be
specified based on local properties of the manifold or the density of
sources at a given point.  The literature on LLE offers no consensus
on how to choose the optimal value of $K$.  Too large a value, and the
manifold is oversampled, destroying the assumption of local linearity,
and leading to an inherently larger dimensional space.  Too low, and
the manifold is undersampled, which leads to a loss of information on
the relationships between neighborhoods and more sensitivity to the
presence of outliers and noise. In this work we take an empirical
approach and try to estimate a series of heuristics that can be
applied to the spectral data. We take a range of values for
$K$ ($10<K<30$) and determine our ability to maximally distinguish
between galaxy populations in the resulting LLE classifications. 
In the case of galaxy spectra, the $K$ is chosen which maximizes the
opening angle between the QSO and emission-line galaxy sequences, measured
between any two of the three leading projection dimensions.  We
find that a value of $K=23$ provides the maximal discrimination but that
values of $K$ from 20 to 27 are comparable.

It should be noted that the final LLE projection is unique only up to
a global translation and rotation.  Because of this, the exact nature of
the projection is highly dependent on the subset of data chosen, 
as well as the above-mentioned input parameters.  
What is invariant, however, is the relationship between each point and
its neighbors.  In particular, an outlying point in one projection
will be an outlier in another projection.  This is what makes LLE
useful: it allows one to easily visualize the \textit{relationships}
between points in a high-dimensional parameter space.

\subsection{Fast LLE: efficient sampling strategies}
\label{sampling}
The LLE algorithm, even when optimized, is computationally expensive.
The main bottleneck is the neighbor search, which, for a single point
is $\mathcal{O}[N D_{in}]$.  Using brute-force for all $N$ points is, at
worst, $\mathcal{O}[N^2 D_{in}]$.  While more advanced tree-based
algorithms can significantly improve on this, 
it remains a computational hurdle, especially for high-dimensional datasets 
\citep[see][for a review of fast nearest-neighbor algorithms]{Berchtold98}.

The second bottleneck is the computation of the global projection,
which involves determining the bottom $D_{out}+1$ eigenvectors of a
$[N\times N]$ sparse matrix (see appendix \ref{finding_y}).  By using
iterative techniques such as Arnoldi/Lanczos Decomposition, it is possible to
determine these eigenvectors without performing a full matrix
diagonalization \citep[see][for one well-tested optimized algorithm]{ARPACK}.

Even fully optimized, it is not feasible to learn the LLE projection
from the full
SDSS spectral sample, which would amount to hundreds of thousands of
points in a 4000-dimensional dataspace.  What is more, a random
subsample of these data will be too sparsely populated in some regions
of space (e.g.\ for strong line emitters that account for $<0.01$\% of
a randomly sample of spectra from the SDSS), and too densely populated
in others (e.g.\ around L$^*$ galaxies). In this case LLE will not
sufficiently probe the entire structure of the manifold.  For this
reason, a strategy is needed to define a subsample that best spans the
entire parameter space, without overpopulating the most probable
neighborhoods.

To do this, we follow \citet{deRidder2002} and appeal to the
fundamental assumption of local linearity on the manifold.  For a
given point \matvec{x_i}, the first step of LLE is to compute the
weights required to construct it from its $K$ neighbors.  The
procedure outlined in section \ref{choosing_parameters} for
determining the intrinsic dimensionality can be used to find the local
reconstruction error: i.e., the percentage of total variance not captured in
the $D_{in}$-dimensional projection.  Points for which this
local projection error is small add very little information to the
overall projection.  They can be discarded without much effect.  
Points for which this reconstruction error is large are not
well-approximated by a linear combination of neighbors, and, therefore,
contain information not present in the local best-fit hyperplane.  If they
are thrown out, the projected manifold will change significantly.

With this in mind, the following strategy was used to reduce the $\sim170,000$
unique spectra to a manageable $9,000$:
\begin{enumerate}
  \item Divide the sample into groups of $2000$ spectra each.
  \item For each group:
    \begin{enumerate}
      \item find the eigenvalues of the local covariance matrix of 
	each point, and create a cumulative distribution based on the
	local reconstruction error.
      \item randomly select 20\% of the points based on this cumulative
	distribution, such that those with the largest reconstruction
	error are preferentially selected.
    \end{enumerate}
  \item merge every 5 subsamples into a new sample of $2000$ spectra
  \item repeat this procedure until the total subsample is of the
    desired size.
\end{enumerate}

\subsection{Applying LLE to new data sets}
\label{new_points}
Once the projection is determined for a given training sample, it is
straightforward to determine the projection of new data onto the
training surface.  
For each new point, the $K$ nearest neighbors are determined within
the training data.  Weights are determined by minimizing the form of
Equation \ref{weight_E}.  These weights are then used within Equation
\ref{proj_E} to determine the projection of the point.  Note that this
second step amounts to nothing more than a weighted sum of the projected
neighbors.  This is a key point in the application of LLE: once a 
projection is defined for a representative training sample, the 
computational cost of the
classification of a set of test-points is largely due to a 
$K$-nearest-neighbor search within the training set.  This can be
done optimally in $\mathcal{O}[D_{in}\log N]$.

Reconstructing unknown data from projected coordinates 
requires the reverse of the above procedure.  Weights are determined
by minimizing the corollary of Equation \ref{weight_E} within the 
projected space, and these weights are used to reconstruct the point
in the original data space.

\section{Discussion}
\label{discussion}

As we have shown in section \ref{LLE_on_spectra}, LLE provides an
efficient and automated way of classifying high dimensional data (in
our case galaxy spectra) that can account for inherent non-linearities
within these data. For spectroscopic observations LLE can jointly
classify spectra based on their emission-line and continuum properties
which makes it well suited to the task of automated classification for large
spectroscopic surveys. In the following section we compare the output
from the LLE classifications with those from standard spectral
classification techniques and discuss how LLE might be adopted for the
next generation of spectroscopic surveys.

\begin{figure}
  \centering
  \includegraphics[width=0.9\textwidth]{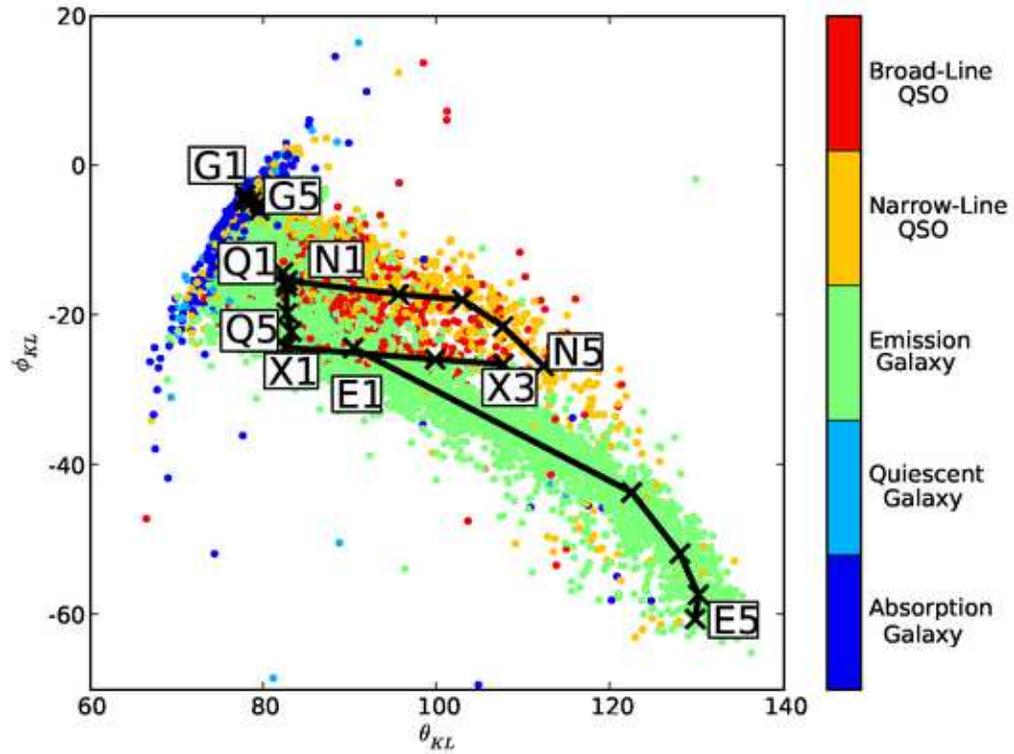}
  \caption{Principal component mixing-angle diagram for the SDSS spectra
    subsample.  The labeled tracks correspond to average spectra plotted
    in figures \ref{em_gal_track}-\ref{quiescent_gal_track}. 
    See caption of Figure \ref{LLEplot} for description of colors.}
  \label{PCA_plot}
\end{figure}

Figure \ref{PCA_plot} shows the PCA mixing-angle plot using the
publicly available \citet{Yip04} SDSS galaxy eigenspectra (see Figure 8
in that work).  The average spectral sequences from Figures
\ref{em_gal_track}-\ref{quiescent_gal_track} are over-plotted for
comparison.  As discussed by \citet{Yip04}, $\phi_{KL}$ correlates
with spectral type \citep[see also][]{Connolly95} 
or star formation rate. It is,
however, not well-suited to distinguishing between various types of
line emission: broad-line QSOs, narrow-line QSOs, and emission-line
galaxies overlap the same region of the projection space.  This is a
reflection of the fact that PCA is sensitive to continuum information
rather than to line emission. Accounting for line emission in a PCA
framework requires extracting the emission lines and applying PCA to
the line equivalent widths \citep{Gyory08}.

\begin{figure}
  \centering
  \includegraphics[width=\textwidth]{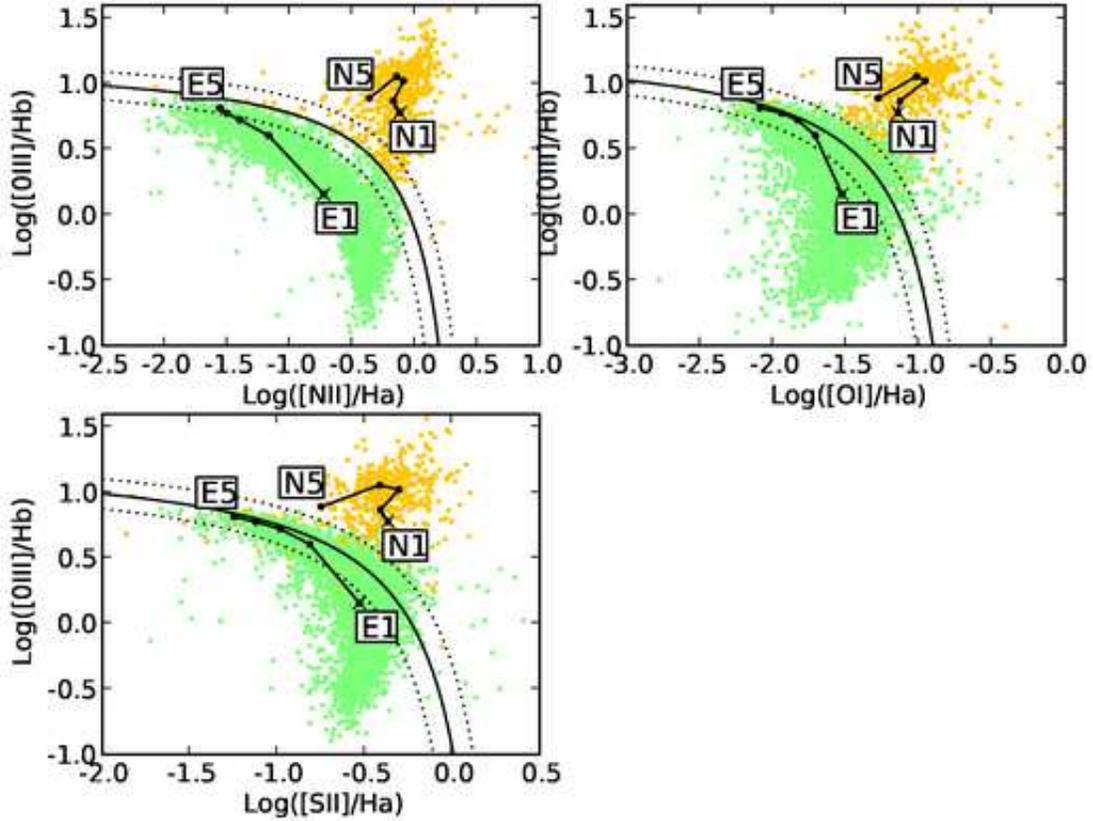}
  \caption{Line-ratio diagram of the objects in the 
    subsample which display $3\sigma$ emission with a width $<1200km$.  
    Points above the cutoff are considered narrow-line QSOs, while points
    below the cutoff are considered emission-line galaxies.
    Note that broad-line QSOs are not meaningfully classified by these
    diagrams, and are omitted. Lines show the boundaries between
    narrow-line QSOs and star-forming galaxies given in \citet{Kewley01}.
    The tracks labeled E and N are the average spectra from Figures 
    \ref{em_gal_track} and \ref{narrow_qso_track}, respectively. }
  \label{BPTplot}
\end{figure}

Figure \ref{BPTplot} shows the typical line-ratios used to distinguish
between star-forming galaxies and narrow-line QSOs.  The dividing
lines between the populations are due to \citet{Kewley01}.  The tracks
on the plot correspond to the average spectra in Figures
\ref{em_gal_track} and \ref{narrow_qso_track} (Note that line-ratio
diagrams do not meaningfully classify broad-line QSOs, which uniformly
occupy the entire space of the line-ratio plots).  \citet{Kewley01}
use stellar synthesis models to show that emission characteristics of
the objects below the dividing line can be attributed to ionization
due to young stellar populations, with the ionization parameter $q$
increasing toward the upper-left.  This gives a physical
interpretation of the emission galaxy (E) sequence of Figure
\ref{em_gal_track}: increasing star formation from E1 to E5.

The region above the dividing line, therefore, is due to ionization
which cannot be attributed to star formation.  \citet{Ho97} show
empirically that this region can be subdivided into transitional
objects near the dividing line, LINERs which lie near N1-N2, and
Seyfert-2 galaxies which lie near N3-N5.  The narrow-line QSO (N)
sequence of Figure \ref{narrow_qso_track} traces this transition.

The above discussion points to the usefulness of LLE for automatic
classification, without requiring the extraction of line parameters
from the spectroscopic data.  As with most classification schemes, the
bulk of the work happens in the training of the LLE. Once the projection 
parameters (Section \ref{choosing_parameters}) and training
subsample (Section \ref{sampling}) are chosen, the resulting projection
provides the basis for a unique, repeatable classification
scheme, with computational cost dominated by a $K$-nearest neighbor
search within the training sample (Section \ref{new_points}).  As such
LLE can provide an accurate initial classification and anomaly
detection step for any automated spectroscopic pipeline, requiring only
the extraction of spectroscopic redshifts.

LLE is not limited to spectroscopic information. In principle, LLE can
be used to visualize any type of data, even non-homogeneous data sets,
e.g.\ a combination of fluxes, redshifts, angular size, etc.  Care
must be taken, though, to make sure that non-homogeneous dimensions
contribute equally to the nearest neighbor search.  This can be
assured through modification of the distance metric for the
nearest-neighbor search such that the distance between point
\matvec{x} and point \matvec{y} are
\begin{equation}
  d_{\matvec{xy}} = \sqrt{ \sum_i w_i(x_i-y_i)^2 }
\end{equation}
where $w_i$ is a weight associated with measurement dimension $i$.
For example, if the measurements in question are fluxes that range 
from $0$ to $1000$
and redshifts that range from $0$ to $1$, $w_i$ could be set to
reflect the parameter range: $w_i = [\max(x_i)-\min(x_i)]^{-1}$.  
This normalization of the data prior to the LLE step is common to 
any dimensional reduction technique, including PCA.

The computational cost of learning the LLE required that we develop a
subsampling technique that preserves the complexity of the underlying
data (i.e.\ as opposed to a random sampling that will be dominated by
the typical sources within a population). Neighborhood-based
subsampling is important in two key areas: it yields a subsample which
fills the entire observational parameter space, and the data are uniformly
sampled throughout.  Furthermore, the local nature of the process
means that subtle nonlinear effects are not obscured in the process.
This subsampling technique, described in Section \ref{sampling}, is
broadly applicable, regardless of the classification scheme being
employed. It could be applied to any problem that requires sampling of
a large data set -- such as spectroscopic follow-up of optical or radio
selected galaxies or spectroscopic observations of galaxies for
training galaxy photometric redshifts.

\section{Conclusions}

Fast classification of large data sets is an increasingly present problem
in astronomy.  With this paper we show Locally Linear Embedding to be a 
powerful tool in the classification SDSS spectra that is sensitive to both 
low-frequency and high-frequency information. As such, is an improvement 
over standard techniques such as PCA/KL decomposition and line-ratio 
diagnostics.  LLE is still in its infancy, and more understanding is 
needed to take into account measurement uncertainties, missing data, 
and undersampled manifolds.  We discuss the limitations of LLE
due to the computational cost of operating on large data sets and 
how these can be addressed through intelligent sampling of the training 
data.  Our sampling technique is based on local information content and 
yields a training subsample which preserves the nonlinear traits of the 
sample as a whole. As a dimensionality reduction technique, 
LLE is both fast and 
accurate, and sensitive to all the information contained in measurement 
data.  Because of this, it has potential to become a useful step in 
efficient classification of large astronomical data sets.

LLE software used in this paper is publicly available at 
http://ssg.astro.washington.edu/software (see appendix \ref{dimred_code}).

\begin{acknowledgments}
We thank S. Anderson, S. Krughoff, A. Becker, and T. Budav{\'a}ri for helpful 
comments and conversations. 
We thank the anonymous referee for several helpful suggestions.
AJC is partially supported through NSF award 0851007.  
JTV is partially supported through NASA award NNX07AH07G 
and NSF award 0851007.
\end{acknowledgments}


\newpage
\appendix

\section{Notational Conventions}
\label{notation}
Throughout this paper, matrices are denoted by bold capital letters 
(\matvec{X}, \matvec{Y}, etc.), vectors are given by bold lowercase 
letters ($\mathbf{x}$, $\mathbf{y}$, etc.), while scalar values are 
in normal-face type ($X$, $y$, $\lambda$, etc.).  Subscripts reference 
the individual elements of matrices, such that $X_{ij}$ is the element 
in the $i$th row and $j$th column of matrix $\mathbf{X}$.  The $i$th 
column vector of matrix $\mathbf{X}$ is given by $\mathbf{x_i}$.  Note 
the potential for confusion with this convention: $(\mathbf{x_j})_i = X_{ij}$. 
To limit this source of confusion, individual elements of a matrix will 
always be referenced via the latter formulation.  \matvec{I} is the 
square identity matrix, such that $I_{ij} = \delta_{ij}$, where 
$\delta_{ij}$ is the Kroniker delta function. \matvec{0} is the zero 
matrix, such that $0_{ij} = 0$ for all $i$ and $j$.  The sizes of these 
two matrices should be clear from the context in which they are used.
The derivations in the following sections are based on those found in
\citet{Roweis2000} and \citet{deRidder2002}, with the addition of a few 
clarifying details.


\section{Determining the Optimal Weights}
\label{finding_w}
The optimal weights $\matvec{w}^{(i)}$ are determined by minimizing the 
reconstruction error given in equation \ref{weight_E}:
\begin{equation}
 \mathcal{E}_1^{(i)}(\matvec{w}^{(i)}) = \Bigg| \matvec{x_i} - 
 \sum_{j=1}^K w_j^{(i)}\matvec{x}_{n_j^{(i)}}\Bigg|^2
\end{equation}
This can be straightforwardly solved in closed form, using the method 
of Lagrangian multipliers.  Consider a neighborhood about the point 
$x_i$, with weights $w_j$ that satisfy 
\begin{equation}
 \label{W_constraint}
 \sum_{j=1}^K w^{(i)}_j = 1 .
\end{equation}
This constraint is useful because it imposes translational invariance to 
the projection.  Using this condition, the above sum can be rewritten 
\begin{equation}
 \mathcal{E}_1^{(i)}(\matvec{w}^{(i)}) = 
 \Bigg|\sum_{j=1}^K w_j^{(i)}\Big(\matvec{x_i} -  
 \matvec{x}_{n_j^{(i)}} \Big)\Bigg|^2.
\end{equation}
Now defining the local covariance matrix
\begin{equation}
  \label{local_covariance_matrix}
  C^{(i)}_{jk} = \Big(\matvec{x_i} -  
  \matvec{x}_{n_j^{(i)}}\Big)^T\Big(\matvec{x_i} -  \matvec{x}_{n_k^{(i)}}\Big)
\end{equation}
We find
\begin{equation}
 \mathcal{E}_1^{(i)}(\matvec{w}^{(i)}) = \sum_{j=1}^K
 \sum_{k=1}^K w_j^{(i)} w_k^{(i)}C^{(i)}_{jk}
\end{equation}
We can minimize this by applying a Lagrange multiplier $\lambda_i$ 
to enforce the constraint given in equation \ref{W_constraint}:
\begin{equation}
 \mathcal{E}_1^{(i)}(\matvec{w}^{(i)}) = 
 \sum_{j=1}^K\sum_{k=1}^K w_j^{(i)} w_k^{(i)}C^{(i)}_{jk} + 
 2\lambda_i \Big(1-\sum_j w_j^{(i)}\Big)
\end{equation}
Minimizing $\mathcal{E}_1^{(i)}(\matvec{w}^{(i)})$ with respect 
to $\matvec{w}^{(i)}$ gives the condition of minimization:
\begin{equation}
 \label{form}
 \matvec{C}^{(i)} \matvec{w}^{(i)} = \lambda_i\matvec{\vec{1}},
\end{equation}
where $\matvec{\vec{1}} = [1,1,...1]^T$.  Defining $\matvec{R}^{(i)}$ 
to be the inverse of the local covariance matrix $\matvec{C}^{(i)}$, we 
are left with
\begin{equation}
 \label{w_def}
 \matvec{w}^{(i)} = \lambda_i \matvec{R}^{(i)} \matvec{\vec{1}}
\end{equation}
Now imposing the constraint (\ref{W_constraint}) leads to
\begin{equation}
 \label{lambda_def}
 \lambda_i = \Big(\sum_{j=1}^N \sum_{k=1}^N R^{(i)}_{jk}\Big)^{-1}.
\end{equation}

Evaluation of $\matvec{w}^{(i)}$ with equations \ref{w_def} and 
\ref{lambda_def} requires explicit inversion of the local covariance 
matrix.  In practice, a more efficient method of calculation is to 
solve a variation of equation \ref{form},
\begin{equation}
  \label{Cw=1}
 \matvec{C}^{(i)}\matvec{w}^{(i)} = [1,1,1...1]^T ,
\end{equation}
and then rescale the resulting $\matvec{w}^{(i)}$ such that the 
constraint in equation \ref{W_constraint} is satisfied.

In general, the local covariance matrix $\matvec{C}^{(i)}$ may be nearly 
singular.  In this case, the weights $\matvec{w}^{(i)}$ are not well defined. 
\citet{deRidder2002} note that the matrix can be regularized based on its
eigenvalues $\matvec{\lambda}^{(i)}$.  Because, in the end, we are doing a 
global projection to a dimension $D_{out}$, we can follow probabilistic
PCA and define 
\begin{equation}
 r = \frac{1}{K-D_{out}}\sum_{j=D_{out}+1}^{K} \lambda_j^{(i)} ,
\end{equation}
i.e., the mean of the eigenvalues of the unused eigenvectors,
and regularize our matrix as 
$\matvec{C}^{(i)} = \matvec{C}^{(i)} + r\matvec{I}$.  In practice, this
would require an explicit eigenvalue decomposition for each local covariance
matrix, which is very computationally expensive.  We can obtain a similar
regularization parameter by noting that 
$\sum \lambda_i = {\rm tr}(\matvec{C}^{(i)})$.
If our $D_{out}$-dimensional projection contains the majority of the variance
(which, under the assumptions of the algorithm, it should), we see that
$r$ will be very small compared to the trace of $\matvec{C}^{(i)}$.
For this paper, the value $r = (10^{-3}){\rm tr}(\matvec{C}^{(i)})$ was
used.


\section{Determining The Optimal Projection}
\label{finding_y}
The projection error is given by equation \ref{proj_E}:
\begin{equation}
 \mathcal{E}_2(\matvec{Y}) = 
 \sum_{i=1}^N \Bigg| \matvec{y_i} - 
 \sum_{j=1}^K w_j^{(i)}\matvec{y}_{n_j^{(i)}}\Bigg|.
\end{equation}
To simplify this, we define a sparse weight matrix \matvec{W} with 
columns populated by the weight vectors $\matvec{w}^{(i)}$, such that
\begin{equation}
 W_{ik} = \left\{
 \begin{array}{ll}
   w_j^{(i)} & \mathrm{if\ }k = n_j^{(i)} \\
   0         & \mathrm{if\ }k \notin \matvec{n}^{(i)}
 \end{array}\right.
\end{equation}  
The projection error can then be rewritten in a more compact form:
\begin{equation}
 \mathcal{E}_2(\matvec{Y}) = \Big| \matvec{Y} - \matvec{Y}\matvec{W}\Big|^2.
\end{equation}
It can be shown that this is equivalent to the quantity
\begin{equation}
 \label{E_Y}
 \mathcal{E}_2(\matvec{Y}) = \mathrm{tr}\Big( \matvec{YMY}^T \Big) ,
\end{equation}
where we have defined
\begin{equation}
 \label{M_def}
 \matvec{M} = (\matvec{I}-\matvec{W})(\matvec{I}-\matvec{W})^T .
\end{equation}  

It is clear that equation \ref{E_Y} is trivially minimized by 
$\matvec{Y} = \matvec{0}$.  In order to find solutions of interest, 
we will impose the constraint
\begin{equation}
 \label{Y_constraint}
 (\matvec{YY}^T) = \matvec{I}.
\end{equation}
That is, the rows of \matvec{Y} are orthonormal.  We can then use 
Lagrangian multipliers $\lambda_i$ to constrain equation \ref{E_Y}:
\begin{equation}
 \mathcal{E}_2(\matvec{Y}) = \mathrm{tr}\Big( \matvec{YMY}^T + 
 \matvec{\Lambda}(\matvec{I}-\matvec{YY}^T)\Big)
\end{equation}
where \matvec{\Lambda} is the diagonal matrix of Lagrangian multipliers, 
with $\Lambda_{ii}=\lambda_i$.  Minimizing this with respect to \matvec{Y} 
gives the condition
\begin{equation}
 \label{Y_eigs_1}
 \matvec{MY}^T - \matvec{Y}^T\matvec{\Lambda} = \matvec{0},
\end{equation}
Denoting the rows of \matvec{Y} with the vectors $\matvec{y}^{(i)}$, 
this can be written in a more familiar form:
\begin{equation}
 \label{Y_eigs_2}
 \matvec{My}^{(i)} - \lambda_i\matvec{y}^{(i)} = \matvec{0}.
\end{equation}
We see that $\lambda_i$ and $\matvec{y}^{(i)}$ are simply the eigenvalues 
and eigenvectors of the symmetric-real matrix \matvec{M}.  Equation 
\ref{Y_eigs_1} can then be rewritten $\matvec{YMY}^T = \matvec{\Lambda}$ 
and our reconstruction error (\ref{E_Y}) is simply
\begin{equation}
 \mathcal{E}_2(\matvec{Y}) = \mathrm{tr}(\matvec{\Lambda}) = \sum_i \lambda_i .
\end{equation}

Evidently, the desired projection is contained in the eigenvectors of 
\matvec{M} corresponding to the smallest eigenvalues.  Because of the 
constraint in equation \ref{W_constraint}, it is clear that 
$\matvec{y}^{(0)} \propto [1,1,1,...1]^T$ is a solution with 
eigenvalue $\lambda_0 = 0$.  This solution amounts to a 
translation in space, and can be neglected.

Thus, we see that the $D_{out}$-dimensional projection \matvec{Y} of a 
data matrix \matvec{X} is given simply by the $(D_{out}+1)$-dimensional 
null-space of matrix 
$\matvec{M} = (\matvec{I}-\matvec{W})(\matvec{I}-\matvec{W})^T$, 
with the lowest eigenvector neglected. 

Fortunately, a full matrix diagonalization is not necessary 
to find only a few extreme eigenvectors.  Iterative methods such
as Arnoldi decomposition or Lanczos factorization can be used
to determine this null-space rather efficiently.  What is more,
though the matrix $\matvec{I}-\matvec{W}$ is a sparse $[N\times N]$ matrix,
with only $K+1$ nonzero entries per column.  Thus, the LLE algorithm
has large potential for optimization in both computational and storage
requirements.

\section{LLE Algorithm}
\label{algorithm}
Here is a description of the basic LLE algorithm, taking into account the
above discussion.  Given $N$ points in $D_{in}$ dimensions, to be
projected to $D_{out}$ dimensions, using the $K$ nearest neighbors
of each point:
\begin{enumerate}
  \item for $i=1\to N$
    \begin{enumerate}
      \item find the $K$ nearest neighbors of point $i$
      \item construct the local covariance matrix, $\matvec{C}^{(i)}$ 
	(eqn. \ref{local_covariance_matrix})
      \item regularize the local covariance matrix by adding a fraction
	of its trace to the diagonal: 
	$\matvec{C}^{(i)} = \matvec{C}^{(i)} + 
	\delta\cdot{\rm tr}(\matvec{C}^{(i)})
	\matvec{I}$.  Choosing $\delta = 10^{-3}$ works 
	well in practice (See appendix \ref{finding_w}.)
      \item determine $\matvec{w}^{(i)}$ by solving equation \ref{Cw=1}.
      \item use $\matvec{w}^{(i)}$ to populate column $i$ of the 
	weight matrix $\matvec{W}$, with
	$W_{ji}=0$ if point $j$ is not a neighbor of point $i$.
    \end{enumerate}
  \item Construct the matrix $\matvec{M} = (\matvec{I}-\matvec{W})
    (\matvec{I}-\matvec{W})^T$
  \item Determine the $D_{out}+1$ eigenvectors of matrix $\matvec{M}$ 
    corresponding to the smallest eigenvalues. The LLE projection is
    given by these, neglecting the lowest (which is merely a translation).
\end{enumerate}


\section{Description of DimReduce code}
\label{dimred_code}
\texttt{DimReduce} is a C++ code which performs LLE and some of its
variants on large datasets.  To handle the computationally intensive
pieces of the algorithm, it includes a basic interface to the
optimized FORTRAN routines in \textit{BLAS}, \textit{LAPACK}, 
and \textit{ARPACK}.

The \texttt{DimReduce} interface works with data saved in the \texttt{FITS}
format, a commonly used binary file format used for astronomical images
and data.  Four variants of LLE are available:
\begin{description}
  \item[LLE:] This is the basic LLE algorithm.  The user must provide input
    data, the number of nearest neighbors, and either the output dimension
    or a desired variance to preserve within each neighborhood.

  \item[HLLE:] This is a variant of LLE which uses a Hessian estimator 
    within each neighborhood which better preserves angles in each 
    neighborhood.  The algorithm is based on \citet{Donoho03}.  Inputs 
    are similar to LLE, with the added restriction that $K$ must be 
    larger than $D_{in}$, the input dimension.

  \item[RLLE1:] This is a robust variant of LLE, based on an algorithm
    outlined in \citet{Chang06}.  To detect outliers, an
    iteratively reweighted PCA is performed on each neighborhood.  
    A reliability score is assigned to each point based on the number of
    neighborhoods in which it appears, as well as its iteratively determined
    contribution within each neighborhood.  The user provides inputs similar
    to those of LLE, as well as an reliability cutoff: 
    a number between zero and one that
    represents the fraction of points believed to be outliers.

  \item[RLLE2:] This is another robust variant of LLE outlined in
    \citet{Chang06}, where outliers are
    handled on a neighborhood by neighborhood basis.  Reliability scores
    are computed as described above, and within each neighborhood the nearest
    $K+R$ nearest neighbors are found, and then the $R$ neighbors with the
    lowest reliability scores are discarded.  The user provides inputs
    similar to those of LLE, as well as the number $R$ of excess neighbors
    to find within each neighborhood.

  \item[LLEsig:] This is a utility routine which can be used in the data
    selection procedure described in section \ref{sampling}.  Given an 
    input matrix, a number of nearest neighbors, and an output dimension,
    this will compute the local linear reconstruction error for each point.
    The result is the local reconstruction error for each point.
\end{description}
In addition, a python utility is provided to plot and examine the results,
using the open source packages \texttt{scipy} and \texttt{matplotlib}.
All of the source code is available on the 
web\footnote{http://ssg.astro.washington.edu/software}, 
along with test data and example scripts.

There are a few possible optimizations to \texttt{DimReduce} which have
not yet been implemented.  First of all, the sparse 
weight matrix is presently stored  in dense format.  
Storage requirements and execution time could be greatly reduced  
by taking advantage of its sparsity.  
Secondly, the $K$-nearest neighbor search
is presently performed using a non-optimized brute-force algorithm.  
Use of a ball-tree, cover-tree, or similar algorithm 
would greatly speed up the performance, especially when 
projecting new test data onto a training surface.
Thirdly, the entire package could be fairly easily restructured to take
advantage of parallel computing capabilities.  Parallelized versions of
\textit{BLAS}, \textit{LAPACK}, and \textit{ARPACK} are all publicly 
available, and the computation of nearest neighbors and construction
of the weight matrix lends itself 
easily to parallelization.  Implementation of these 
improvements, along with the use of the data sampling strategies outlined 
in section \ref{sampling} give LLE the possibility of being used for
automated classification of objects in a large variety of contexts.
 
\bibliography{SDSS_LLE}
\end{document}